\documentclass[prd,preprint,tightenlines,superscriptaddress,floatfix,showpacs,preprintnumbers,nofootinbib,eqsecnum]{revtex4}

 \usepackage[dvips,final]{graphicx}
     \usepackage{epsfig}
      \usepackage{bm}

\begin{document}

\thispagestyle{empty} \preprint{\hbox{}} \vspace*{-10mm}

\title{Exclusive muon-pair productions in ultrarelativistic heavy-ion collisions: Realistic nucleus charge form factor and differential distributions}

\author{M.~K{\l}usek-Gawenda}
\email{mariola.klusek@ifj.edu.pl}

\affiliation{Institute of Nuclear Physics PAN, PL-31-342 Cracow, Poland}

\author{A.~Szczurek}
\email{antoni.szczurek@ifj.edu.pl}

\affiliation{Institute of Nuclear Physics PAN, PL-31-342 Cracow, Poland} 
\affiliation{University of Rzesz\'ow, PL-35-959 Rzesz\'ow, Poland}

\date{\today}

\begin{abstract}
The cross sections for exclusive muon pair production 
in nucleus - nucleus collisions are calculated and
several differential distributions are shown. 
Realistic (Fourier transform of charge density)
charge form factors of nuclei are used and the corresponding
results are compared with the cross sections calculated
with monopole form factor often used in the literature
and discussed recently in the context of higher-order
QED corrections. 
Absorption effects are discussed and quantified.
The cross sections obtained with realistic 
form factors are significantly smaller than those obtained
with the monopole form factor. The effect is bigger for
large muon rapidities and/or large muon transverse momenta.
The predictions for the STAR and PHENIX collaboration measurements 
at RHIC as well as the ALICE and CMS collaborations at LHC are presented. 
\end{abstract}


\maketitle

\section{Introduction}

\begin{figure}[!h]    
\begin{minipage}[t]{0.4\textwidth}
\centering
\includegraphics[width=0.9\textwidth]{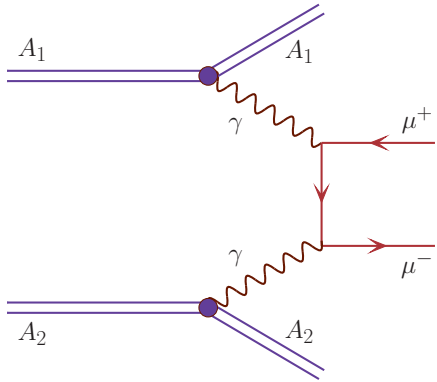}
\end{minipage}
   \caption{\label{fig:diagram_Born}
   \small (Color online) The Born diagram for the exclusive dimuon production.
}
\end{figure}

In Fig.\ref{fig:diagram_Born} we show the basic QED mechanism
of the exclusive production of muon pairs. The shaded circles
represent the coupling of photons to large-size objects -- 
nuclei. In the momentum space this is done in terms of 
electromagnetic form factors of nuclei. In the case of scalar
nuclei there is only one form factor -- the charge form factor
of the nucleus.

It was recognized long ago that the production rate of 
leptons in ultrarelativistic heavy ion collisions is enhanced
considerably by the coherent effects and large charge
of colliding ions \cite{Budnev}. Many results have been
presented in the literature since then 
(for reviews of the field see e.g. 
\cite{BHTSK02,Baltz_review}).
Recently, there was a growing theoretical interest in
estimating higher-order QED corrections 
\cite{Nikolaev,Serbo,Baltz1,Baltz2}.
In most of the practical calculations of exclusive dilepton production a simple monopole
charge form factor of the nucleus was used. While it may 
be sufficient for estimating the total cross section, 
it may be not sufficient for calculations of the differential cross sections. 
The importance of including realistic charge form factors 
was discussed recently for exclusive production of 
pairs of $\rho^0$ mesons \cite{KSS09}. 

Most of the existing calculations concentrated on
total cross section, interesting theoretical quantity,
which cannot be, however, measured in practice, neither at RHIC 
nor at LHC. The experiments running at RHIC and those planned at LHC
demand severe cuts on lepton transverse momenta
or on their rapidities. 

It is the aim of the present analysis to make realistic 
estimates of the cross sections including the experimental 
cuts.
We shall compare the results obtained with monopole form 
factor used in the literature and the results obtained with realistic
form factor being Fourier transform of the charge density
of the nucleus. We shall perform the calculation in the
equivalent photon approximation (EPA) in the impact parameter
space as well as in the momentum space. 
While the impact parameter EPA allows to include easily 
absorption effects due to the size of colliding nuclei, 
the momentum space approach allows to study easily several 
differential distributions. In our calculation we shall include
experimental limitations of the STAR and PHENIX detectors
at RHIC and those of the ALICE and CMS detectors at LHC.

\section{Formalism}

\subsection{Charge form factor of nuclei}

The charge distribution in nuclei is usually obtained from
elastic scattering of electrons from nuclei \cite{Barrett_Jackson}.
The charge distribution obtained from these experiments is 
often parametrized with the help of two--parameter Fermi model 
\cite{nuclear_density}:
\begin{equation}
 \rho \left( r \right) = \rho_0 \left( 1 + exp \left( \frac{r-c}{a} \right) \right)^{-1},
\end{equation}
where $c$ is the radius of the nucleus and $a$ is the so-called diffiusness parameter of the charge density. 

\begin{figure}[!h]    
\begin{minipage}[t]{0.46\textwidth}
\centering
\includegraphics[width=1\textwidth]{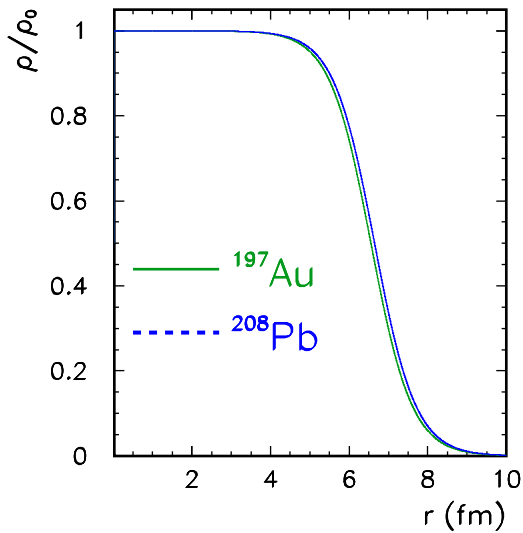}
\end{minipage}
   \caption{\label{fig:density}
   \small
(Color online) The ratio of the charge distibution ($\rho$) to the density in the center of nucleus ($\rho_0$).
}
\end{figure}
Fig.~\ref{fig:density} shows the charge density normalizationed to 
unity at $r=0$. The correct normalization is:
$\rho_{Au}(0) = \frac{0.1694}{A} fm^{-3}$ for Au nucleus and 
$\rho_{Pb}(0) = \frac{0.1604}{A} fm^{-3}$ for Pb nucleus. 

\begin{figure}[!h]    
\begin{minipage}[t]{0.46\textwidth}
\centering
\includegraphics[width=1\textwidth]{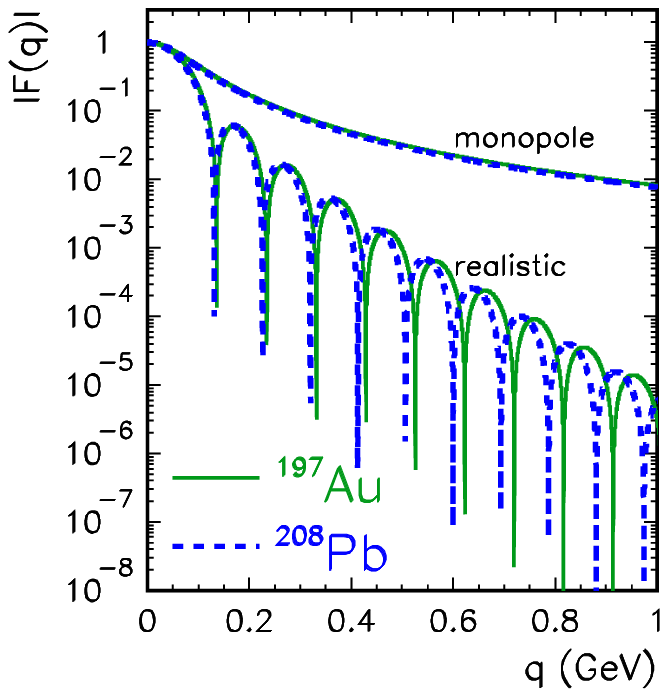}
    \caption{\label{fig:ff_all}
    \small
(Color online) The moduli of the charge form factor $F_{em} \left ( q \right)$ of the $^{197}Au$ and $^{208}Pb$ nuclei for realistic charge distributions. For comparison we show the monopole form factor for the same nuclei.}
\end{minipage}
\hspace{0.03\textwidth}
\begin{minipage}[t]{0.46\textwidth}		
\centering
\includegraphics[width=1\textwidth]{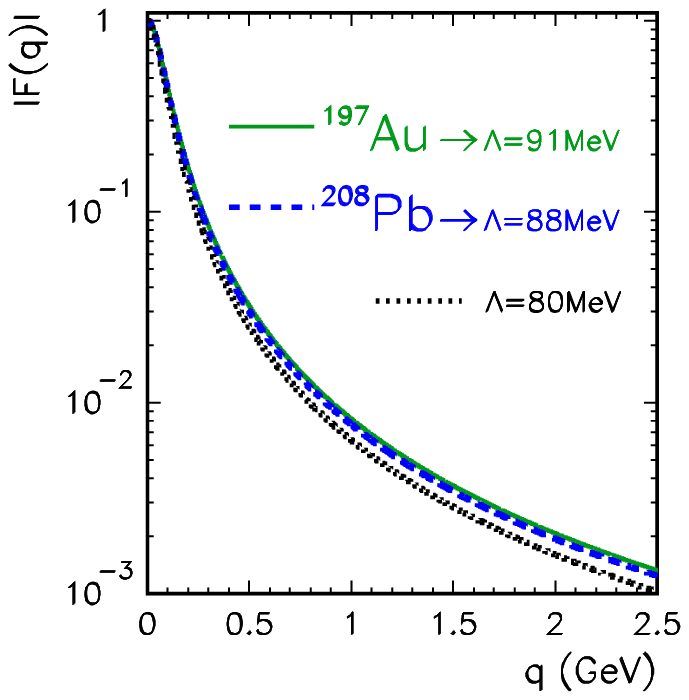}
   \caption{\label{fig:ff_mon}
   \small
(Color online) The monopole form factor for the values of $\Lambda$ reproducing charge radius of $^{197}Au$ and $^{208}Pb$ nuclei and for comparison for $ \Lambda=0.08$ GeV often used in the literature.
}
\end{minipage}
\end{figure}

The form factor ($F$) is the Fourier transform of the charge distribution \cite{Barrett_Jackson}. If $\rho \left( r \right)$ is 
spherically symmetric then the form factor is a function 
of photon virtuality ($q$) only:
\begin{equation}
 F(q) = \int \frac{4 \pi}{q} \rho \left( r \right) sin \left( qr \right) r dr = 1 - \frac{q^2 \langle r^2 \rangle}{3!} + \frac{q^4 \langle r^4 \rangle}{5!} \ldots  \; .
\end{equation}
Fig.~\ref{fig:ff_all} shows the moduli of the form factor as a 
function of momentum transfer. The results are depicted for the gold 
(solid line) and lead (dashed line) nuclei for realistic charge 
distribution. The realistic form factor is 
obtained as a Fourier transform of the realistic charge density 
which we take from the literature \cite{Barrett_Jackson}. 
Here one can see many oscillations characteristic for relatively 
sharp edge of the nucleus. For comparison we show the monopole form factor 
often used in the literature. The two form factors 
coincide only in a very limited range of $q$
and with larger value of $q$ the difference between them 
becomes larger and larger.\\

The monopole form factor \cite{HTB94} given by the simple formula:
\begin{equation}
 F(q^2) = \frac{\Lambda^2}{\Lambda^2 + q^2}
\end{equation}
leads to a simplification of many formulae for photon-photon
collisions.
In our calculation  $\Lambda$ is adjusted to reproduce 
the root mean square radius of a nucleus ($\Lambda = \sqrt{\frac{6}{<r^2>}}$)
with the help of experimental data \cite{nuclear_density}:
\begin{itemize}
 \item for $^{197}Au$: $<r^2>^{1/2}= 5.3 \, \Rightarrow \, \Lambda= 0.091 \, \mbox{GeV}$,
 \item for $^{208}Pb$: $<r^2>^{1/2}= 5.5016 \, \Rightarrow \, \Lambda= 0.088 \, \mbox{GeV}$.
\end{itemize}
Different values of $\Lambda$ are used in the literature, ranging from $80$ to $90$ MeV.
Fig.~\ref{fig:ff_mon} shows the monopole form factor with 
$\Lambda$ adjusted to reproduce the rms radius of the charge distribution. 

\subsection{Equivalent Photon Approximation}

\begin{figure}[!h]    
\begin{minipage}[t]{0.33\textwidth}
\centering
\includegraphics[width=1.1\textwidth]{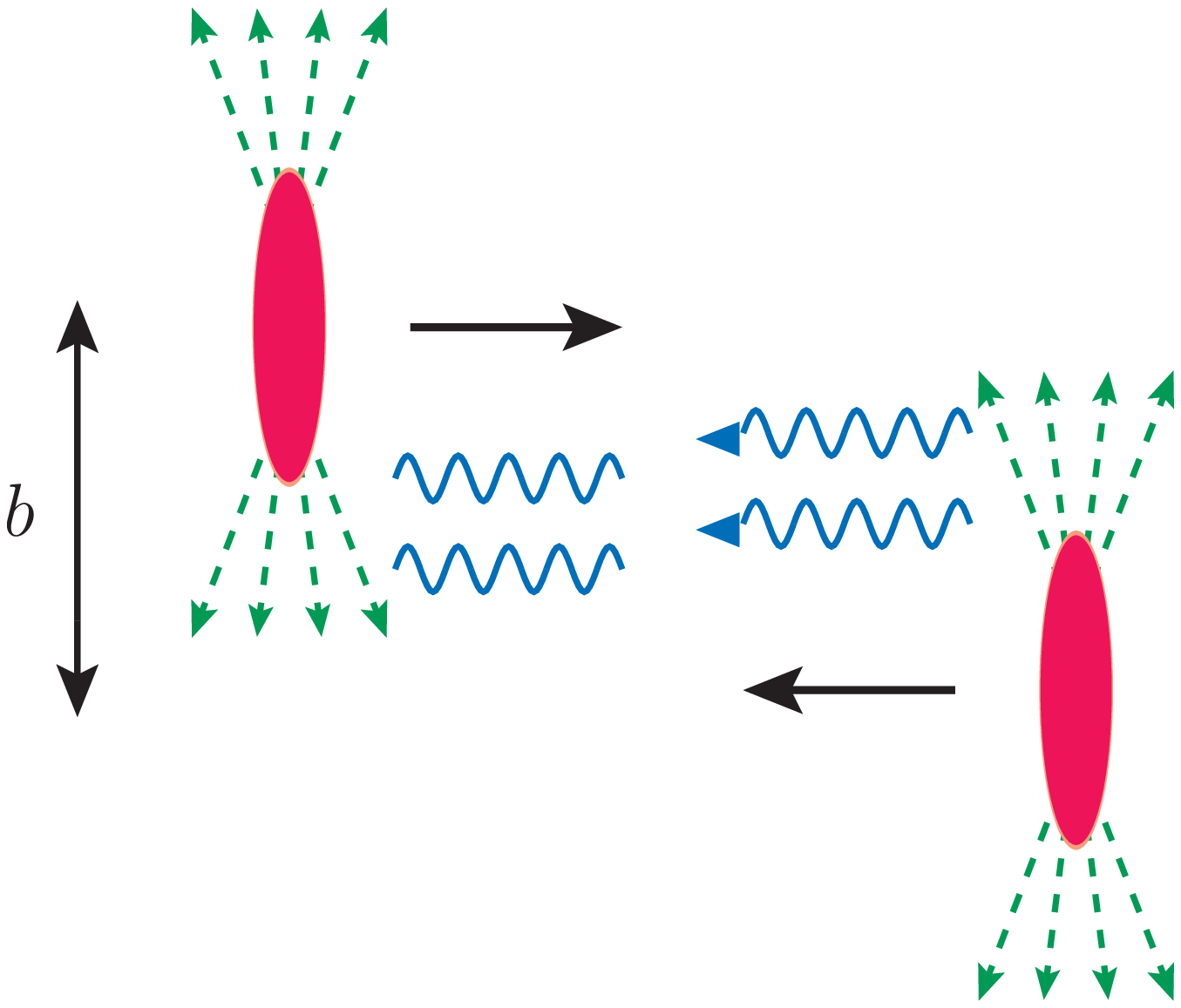}
    \caption{\label{fig:EPA}
    \small 
    (Color online) Equivalent photon approximation.}
\end{minipage}
\hspace{0.03\textwidth}
\begin{minipage}[t]{0.4\textwidth}		
\centering
\includegraphics[width=0.9\textwidth]{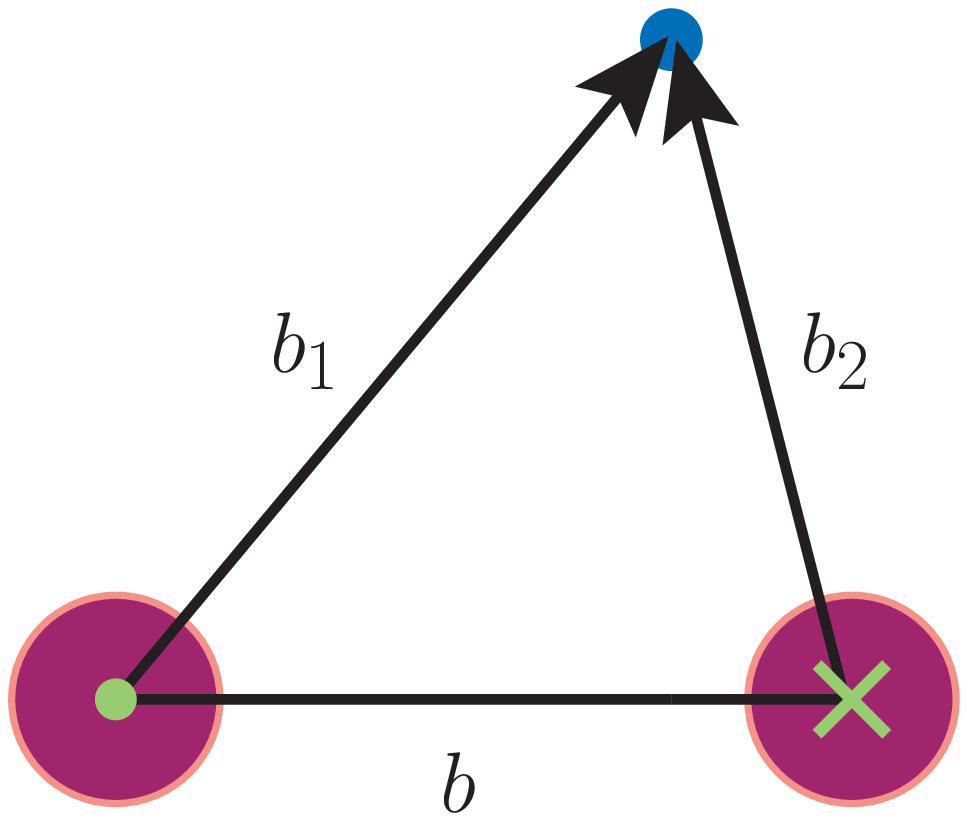}
   \caption{\label{fig:impact_parameter}
   \small
(Color online) The quantities used in the impact parameter calculation.
}
\end{minipage}
\end{figure}

The equivalent photon approximation is the standard semi--classical 
alternative to the Feynman rules for calculating cross sections 
of electromagnetic interactions \cite{Jackson}. 
This is illustrated in Fig.~\ref{fig:EPA} where we can 
see a fast moving nucleus with the charge $Ze$. 
Due to the coherent action of all the protons in the nucleus, 
the electromagnetic field surrounding (the dashed lines are lines 
of electric force for a particles in motion) 
the ions is very strong. This field can be viewed as a cloud of 
virtual photons. 
These photons are often considered as real. They are called 
''equivalent'' or ''quasireal photons''. In the collision of 
two ions, these quasireal photons can collide with each other 
or with the other nucleus. So the strong 
electromagnetic field is used as a source of photons to induce 
electromagnetic reactions on the second ion. 
We consider very peripheral collisions. It means 
that the distance between nuclei is bigger 
than the sum of the radii of the two nuclei ($b>R_1+R_2\cong14fm$). 
Fig.~\ref{fig:impact_parameter} explains the quantities 
used in the impact parameter calculation. We can see a view in 
the plane perpendicular to the direction of motion of the two ions.
In order to calculate the cross section of a process it is 
convenient to introduce a new kinematic variable: 
$x = \frac{\omega}{E_A}$, where $\omega$ is the energy of the photon and the energy of the nucleus 
$E_A=\gamma A m_{proton} = \gamma M_A$, where $M_A$ is the mass of the nucleus 
and $\gamma$ is the Lorentz factor.

The total cross section can be calculated by the convolution:
\begin{eqnarray}
 \sigma\left(AA \rightarrow \mu^+ \mu^- AA ;s_{AA}\right) = 
\int   {\hat \sigma} \left(\gamma\gamma\rightarrow \mu^+\mu^-;
W_{\gamma \gamma} = \sqrt{x_1 x_2 s_{AA}} \right) \, 
{\rm d}n_{\gamma\gamma}\left(x_1,x_2,{\bf b}\right)  .
\end{eqnarray}
The effective photon fluxes can be expressed through the electric 
fields generated by the nuclei:
\begin{eqnarray}
 {\rm d}n_{\gamma\gamma}\left(x_1,x_2,{\bf b}\right) = &  & \frac{1}{\pi} {\rm d^2}{\bf b}_1 |{\bf E}\left(x_1,{\bf b}_1 \right)|^2 \frac{1}{\pi} {\rm d^2}{\bf b}_2  |{\bf E}\left(x_2,{\bf b}_2 \right)|^2   \nonumber \\
 & \times & S^2_{abs}\left({\bf b} \right) \delta^{\left(2\right)} \left({\bf b}-{\bf b}_1 + {\bf b}_2 \right)   \frac{{\rm d}x_1}{x_1} \frac{{\rm d}x_2}{x_2}. 
\end{eqnarray}
The presence of the absorption factor $S^2_{abs}\left({\bf b} \right)$ 
assures that we consider only peripheral collisions, when the nuclei 
do not undergo nuclear breakup. In the first approximation this can be
expressed as:  
\begin{eqnarray}
 S^2_{abs}\left({\bf b} \right)=\theta \left({\bf b}-2R_A \right) = \theta \left(|{\bf b}_1-{\bf b}_2|-2R_A \right) \; .
\end{eqnarray}
Thus in the present case, we concentrate on processes with final 
nuclei in the ground state. 
The electric field strength can be expressed through the charge 
form factor of the nucleus:
\begin{eqnarray}
 {\bf E}\left(x,{\bf b}\right) = Z \sqrt{4\pi \alpha_{em}} \int \frac{{\rm d^2}{\bf q}} {\left(2\pi^2\right)} e^{-i{\bf bq}}\frac{{\bf q}}{{\bf q}^2+x^2M_A^2}F_{em}\left({\bf q}^2+x^2M_A^2\right).
\end{eqnarray}
Next we can benefit from the following formal substitution:
\begin{eqnarray}
\frac{1}{\pi} \int {\rm d^2}{\bf b} |{\bf E}\left(x,{\bf b}\right)|^2 
= \int {\rm d^2}{\bf b} N \left( \omega, {\bf b} \right) 
\equiv n\left(\omega \right) \; 
\end{eqnarray}
by introducting effective photon fluxes which depend on energy of 
the quasireal photon $\omega$ and the distance from the nucleus 
in the plane perpendicular to the nucleus motion $\overrightarrow{b}$. 
Then, the luminosity function can be expressed in term of the photon flux 
factors attributed to each of the nuclei
\begin{eqnarray}
 {\rm d}n_{\gamma\gamma}\left(\omega_1,\omega_2,{\bf b}\right) & = & \int  \theta \left(|{\bf b}_1-{\bf b}_2|-2R_A \right) N \left(\omega_1,{\bf b}_1 \right)    N\left(\omega_2,{\bf b}_2 \right) {\rm d^2}{\bf b}_1 {\rm d^2}{\bf b}_2 {\rm d}\omega_1 {\rm d}\omega_2. 
\end{eqnarray}
The total cross section for the $AA \rightarrow \mu^+ \mu^- AA $ 
process can be factorized into the equivalent photons spectra
( $n \left( \omega \right)$ ) and the 
$\gamma \gamma \to \mu^+ \mu^-$ subprocess cross section as 
(see e.g.\cite{BF91}):
\begin{eqnarray}
 \sigma\left(AA \rightarrow \mu^+ \mu^- AA ; s_{AA}\right)  & = &  
\int   
{\hat \sigma}\left(\gamma\gamma\rightarrow \mu^+ \mu^-; 
W_{\gamma \gamma}  \right) \, 
\theta \left(|{\bf b}_1-{\bf b}_2|-2R_A \right)  \nonumber \\
 &\times & N\left(\omega_1,{\bf b}_1 \right)
N\left(\omega_2,{\bf b}_2 \right)
  {\rm d^2}{\bf b}_1 
 {\rm d^2}{\bf b}_2 {\rm d}\omega_1 {\rm d}\omega_2 \; ,
\label{eq.tot_cross_section}
\end{eqnarray}
where $W_{\gamma \gamma}=\sqrt{4 \omega_1 \omega_2}$ is energy in the $\gamma \gamma$ subsystem.
Eq.~(\ref{eq.tot_cross_section}) is a generalization of the simple 
parton model formula (see e.g.\cite{BHTSK02}):
\begin{equation}
\sigma \left( AA \rightarrow \mu^+ \mu^- AA  \right) 
= \int {\hat \sigma} \left(\gamma\gamma\rightarrow \mu^+ \mu^-; 
\sqrt{4 \omega_1 \omega_2} \right) n\left( \omega_1 \right) 
 n\left( \omega_2 \right) {\rm d} \omega_1
{\rm d} \omega_2  \; .
\end{equation}
Additionally, we define 
$ Y=\frac{1}{2} \left( y_{\mu^+}+y_{\mu^-}\right)$, rapidity of 
the outgoing dimuon system which is produced in the photon--photon 
collision. 
Performing the following transformations:
\begin{equation}
\omega_1 = \frac{W_{\gamma \gamma}}{2}e^Y, \qquad \omega_2 = \frac{W_{\gamma \gamma}}{2}e^{-Y} \; ,
\label{eq:omega}
\end{equation}
\begin{equation}
{\rm d}\omega_1 {\rm d}\omega_2 = \frac{W_{\gamma \gamma}}{2} {\rm d}W_{\gamma \gamma} {\rm d} Y \; ,
\label{eq:transf}
\end{equation}
\begin{equation}
{\rm d} \omega_1 {\rm d} \omega_2 \to {\rm d} W_{\gamma \gamma} {\rm d} Y \mbox{ where } \left|\frac{\partial \left( \omega_1, \omega_2 \right) }{\partial \left( W_{\gamma \gamma}, Y \right)}  \right| = \frac{W_{\gamma \gamma }}{2}
\; ,
\label{eq:transf_jac}
\end{equation}
formula (\ref{eq.tot_cross_section}) can be rewritten as:
\begin{eqnarray}
 \sigma\left(AA \rightarrow  \mu^+\mu^- AA ; s_{AA}\right)  = 
 \int   {\hat \sigma}\left(\gamma\gamma\rightarrow \mu^+ \mu^-; W_{\gamma \gamma}  \right) \theta \left(|{\bf b}_1-{\bf b}_2|-2R_A \right) & & \nonumber \\
 \times N\left(\omega_1,{\bf b}_1 \right) 
 N\left(\omega_2,{\bf b}_2 \right)
   \frac{W_{\gamma \gamma}}{2} {\rm d^2}{\bf b}_1  
 {\rm d^2}{\bf b}_2   
 {\rm d}W_{\gamma \gamma} {\rm d} Y
& \; & .
 \label{eq.tot_cross_section_WY}
\end{eqnarray}
Finally, the cross section can be expressed as the five-fold 
integral:
\begin{eqnarray}
 \sigma \left(AA  \rightarrow   \mu^+\mu^- AA ; s_{AA}\right)  = 
 \int  {\hat \sigma}\left(\gamma\gamma\rightarrow \mu^+ \mu^-; W_{\gamma \gamma}  \right) \theta \left(|{\bf b}_1-{\bf b}_2|-2R_A \right)& & \nonumber \\ 
   \times   N \left(\omega_1,{\bf b}_1 \right) N\left(\omega_2,{\bf b}_2 \right)2 \pi b_m \, {\rm d} b_m \, {\rm d} \overline{b}_x \, {\rm d} \overline{b}_y \frac{W_{\gamma \gamma}}{2} {\rm d}W_{\gamma \gamma} {\rm d} Y & \, & , 
\label{eq.tot_cross_section_our}
\end{eqnarray}
where $\overline{b}_x \equiv (b_{1x}+b_{2x})/2$,
      $\overline{b}_y \equiv (b_{1y}+b_{2y})/2$ and
$\vec{b}_m = \vec{b}_1 - \vec{b}_2$ have been introduced.
This formula is used to calculate the total cross section
for the $A A \to A A \mu^+ \mu^-$ reaction as well as the distributions
in $b = b_m$, $W_{\gamma \gamma} = M_{\mu^+ \mu^-}$ and
$Y(\mu^+ \mu^-)$.

Different forms of form factors are used in the literature. 
We compare the equivalent photon spectra 
for an extended charge distribution (realistic case) 
to the monopole case. 
The dependence of the photon flux on the charge 
form factors can be found in \cite{BHTSK02}: 
\begin{equation}
 N \left( \omega,b \right) = \frac{Z^2 \alpha_{em}}{\pi^2} 
 \frac{1}{b^2 \omega} 
 \left( \int u^2 J_1 \left( u \right) F \left( \sqrt{\frac{ \left(\frac{b \omega}{\gamma} \right)^2 +u^2}{b^2}} \right) 
 \frac{1}{ \left( \frac{b \omega}{\gamma} \right)^2 + u^2} du \right)^2,
\label{basic_EPA_flux}
\end{equation}
where $J_1$ is the Bessel function of the first kind and $q$ is 
the four-momentum of the quasireal photon. 
The calculations with the help of realistic form factor are rather 
laborious, so often a simpler monopole form factor
is used \cite{HTB94}. Introducing monopole form factor to
(\ref{basic_EPA_flux}) one gets:
\begin{equation}
 N \left( \omega,b \right) = \frac{Z^2 \alpha_{em}}{\pi^2} \frac{1}{\omega} 
 \left(  \frac{\omega}{\gamma} 
 K_1 \left(  \frac{b \omega}{\gamma} \right) - 
 \sqrt{\frac{\omega^2}{\gamma^2} + \Lambda^2} \;
 K_1 \left(  b \sqrt{\frac{\omega^2}{\gamma^2}+\Lambda^2} \right) \right)^2 \; ,
\label{monopole_EPA_flux}
\end{equation}
where $K_1$ is the modified Bessel function of the second kind.

\begin{figure}[!h]    
\begin{minipage}[t]{0.46\textwidth}
\centering
\includegraphics[width=1\textwidth]{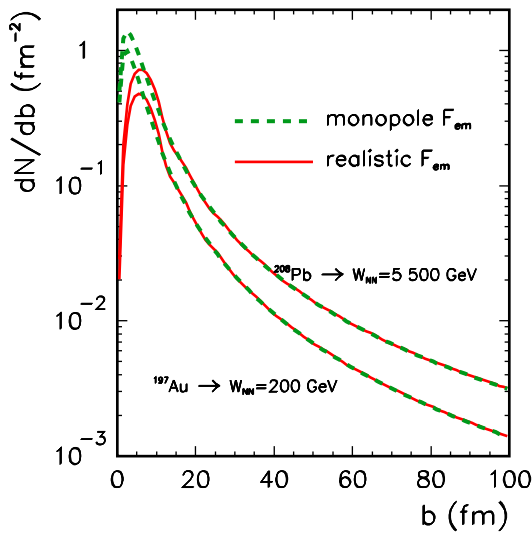}
\end{minipage}
   \caption{\label{fig:dflux}
   \small (Color online) The equivalent photon number as a function of impact parameter (integrated over $\omega$), see Eq.~(\ref{N_b}).
}
\end{figure}

\begin{figure}[!h]    
\begin{minipage}[t]{0.46\textwidth}
\centering
\includegraphics[width=1\textwidth]{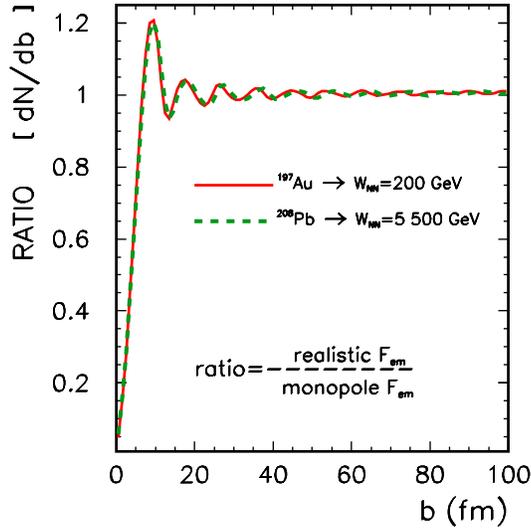}
\end{minipage}
   \caption{\label{fig:dflux_ratio}
   \small (Color online) The ratio of the flux factor obtained with realistic 
charge distribution to that with the monopole form factor 
as a function of impact parameter.
}
\end{figure}
\begin{figure}[!h]    
\begin{minipage}[t]{0.46\textwidth}
\centering
\includegraphics[width=1\textwidth]{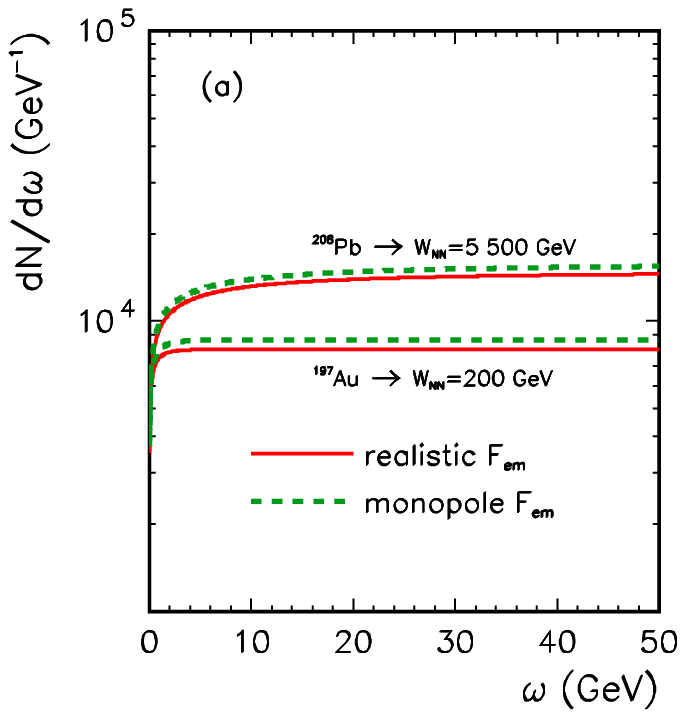}
\end{minipage}
\hspace{0.03\textwidth}
\begin{minipage}[t]{0.46\textwidth}
\centering
\includegraphics[width=1\textwidth]{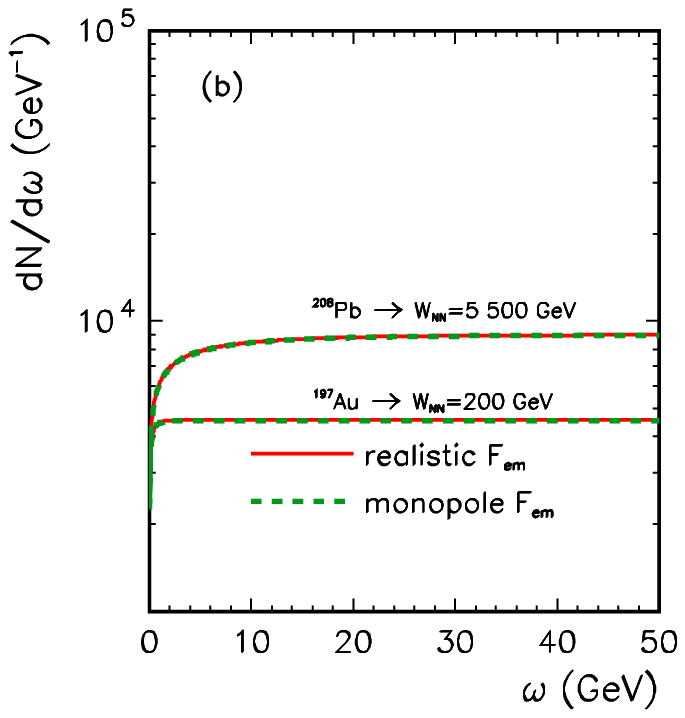}
\end{minipage}
   \caption{\label{fig:flux_omega}
   \small (Color online) The equivalent photon number $n \left( \omega \right)$, see Eq.~(\ref{N_omega}). Left panel: $b\in(0,100)$ fm, right panel: $b\in(14,100)$ fm.
   }
\end{figure}
%
Fig.~\ref{fig:dflux} shows the distribution of the equivalent photon 
number as a function of the impact parameter
\begin{equation}
N \left( b \right) = \int N \left( \omega, \, b \right) d \omega.
\label{N_b}
\end{equation}
We present the results 
for gold and lead nuclei, for realistic and monopole form factors. 
Here we do not impose any sharp cutoff on 
the impact parameter. 
One can see that for small $b$ the flux factor with monopole form factor is bigger. 
For large $b$ the results obtained with the help of realistic and 
monopole form factors are almost the same. 

In addition in Fig.~\ref{fig:dflux_ratio} we show the ratio of 
equivalent photon fluxes obtained with the help of realistic form 
factor to that for the monopole form factor. 
The oscillations in $b$ are due to step-like distribution 
of the charge in the nucleus. 
The results for lower ($\sqrt{s}_{NN}=200$ GeV) and 
higher ($\sqrt{s}_{NN}=5.5$ TeV) energies are almost the same.

Fig.~\ref{fig:flux_omega} shows
\begin{equation}
N \left( \omega \right) = \int 2 \pi b N \left( \omega, \, b \right) db.
\label{N_omega}
\end{equation}
Here we consider the integral over full range of the impact 
parameter (left panel) and for $b>2R_A$ (right panel). 
One can see that the difference between monopole and 
realistic form factor for both gold and lead nuclei is not significant.
The quantity shown depends rather weakly on the photon energy.

\begin{figure}[!h]    
\begin{minipage}[t]{0.46\textwidth}
\centering
\includegraphics[width=1\textwidth]{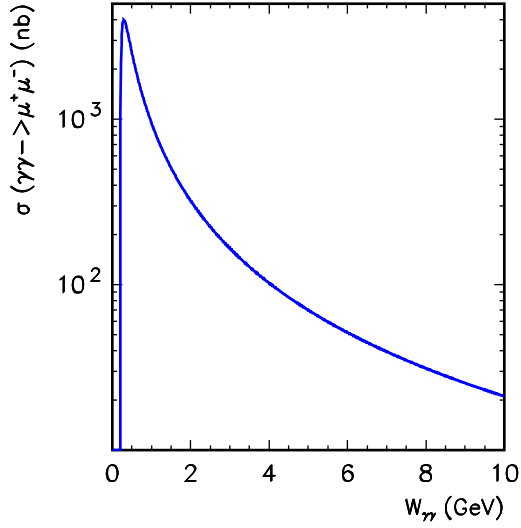}
\end{minipage}
   \caption{\label{fig:el_cross_section_mu}
   \small (Color online) The elementary cross section for the 
$\gamma \gamma \to \mu^+ \mu^-$ reaction as a function of the photon-photon energy.
}
\end{figure}

In Fig.~\ref{fig:el_cross_section_mu} we show the energy dependence
of the elementary $\gamma \gamma \to \mu^+ \mu^-$ cross section 
used in our EPA calculations~\cite{Budnev}:
\begin{eqnarray}
  \sigma  \left( \gamma \gamma \to \mu^+  \mu^- \right) 
&  = & \frac{4 \pi \alpha_{em}^2}{W_{\gamma \gamma}^2}  \\
 & \times &
 \left\{ 2  \ln \left[ \frac{W_{\gamma \gamma}}{2m_{\mu}} 
\left( 1 +  v \right) \right]
  \left( 1+ \frac{4 m_{\mu}^2 W_{\gamma \gamma}^2 -  8 m_{\mu}^4}
{W_{\gamma \gamma}^4} \right)   
- \left( 1 + \frac{4m_{\mu}^2 W_{\gamma \gamma}^2}{W_{\gamma \gamma}^4} \right) v \right\}\, , \nonumber
\label{eq:gg_mumu}
\end{eqnarray}
where
\begin{equation}
v = \sqrt{1-\frac{4m_{\mu}^2}{W_{\gamma \gamma}^2}} \; .
\end{equation}
This formula is often called the Breit-Wheeler formula.

\subsection{Momentum space calculation}

\begin{figure}[!h]    
\centering
\includegraphics[width=0.6\textwidth]{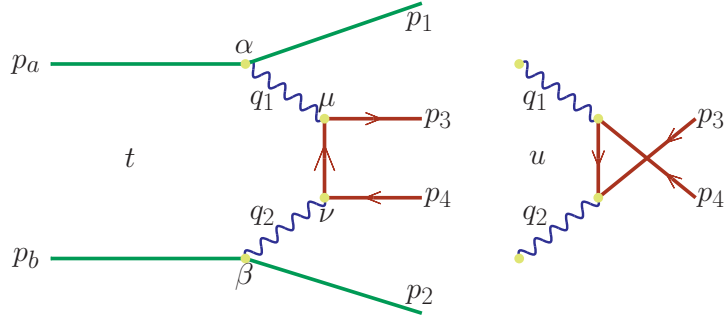}
   \caption{\label{fig:momentum_space}
   \small (Color online) Amplitude of the considered process. On the left one can see the $t$-channel amplitude and on the right - the $u$-channel amplitude.
}
\end{figure}

We consider a genuine $2 \to 4$ reaction 
(see Fig.~\ref{fig:momentum_space}) with four-momenta 
$p_a+p_b \to p_1 +p_2+p_3+p_4$.
In the momentum space approach the cross section for 
the production of a pair of particles can be written as:
\begin{eqnarray}
\sigma & = & \int  \frac{1}{2s} \overline{ | \mathcal{M} | ^2} \left( 2 \pi \right)^4 \delta^4 \left( p_a+p_b-p_1-p_2-p_3-p_4 \right) \nonumber \\
      & \times & \frac{d^3 p_1}{\left( 2 \pi \right)^3 2E_1} \frac{d^3 p_2}{\left( 2 \pi \right)^3 2E_2} \frac{d^3 p_3}{\left( 2 \pi \right)^3 2E_3} \frac{d^3 p_4}{\left( 2 \pi \right)^3 2E_4} \; .
\label{mom_space_1}
\end{eqnarray}
Using
\begin{equation}
\frac{d^3p_i}{E_i}=dy_i \, d^2p_{it}= dy_i p_{it}dp_{it} d\phi_i
\; 
\end{equation} 
Eq.~(\ref{mom_space_1}) can be rewritten as:
\begin{eqnarray}
\sigma & = & \int  \frac{1}{2s} \overline{ | \mathcal{M} | ^2}  \delta^4 \left( p_a+p_b-p_1-p_2-p_3-p_4 \right) \frac{1}{\left( 2 \pi \right)^8} \frac{1}{2^4} \nonumber \\
      & \times &  \left( dy_1 p_{1t} dp_{1t} d \phi_1 \right) \left( dy_2 p_{2t} dp_{2t} d \phi_2 \right) \left( dy_3 d^2p_{3t} \right) \left( dy_4 d^2p_{4t} \right) \; .
      \label{mom_space_2}
\end{eqnarray}
In the above formula 
$p_{it}$ are transverse momenta of outgoing nuclei and leptons, 
$\phi_1$, $\phi_2$ are azimuthal angles of outgoing nuclei. 
Additionally, we introduce a new auxiliary quantity
\begin{equation}
{\bf p}_m = {\bf p}_{3t} - {\bf p}_{4t}
\end{equation}
and benefitting from 4-dimensional Dirac delta function properties, 
Eq.~(\ref{mom_space_2}) can be written as:
\begin{eqnarray}
\sigma & = & \int  \frac{1}{2s} \overline{ | \mathcal{M} | ^2}  \delta \left( E_a + E_b - E_1 - E_2 - E_3 - E_4 \right) \delta^3 \left( p_{1z} + p_{2z} + p_{3z} + p_{4z} \right) \frac{1}{\left( 2 \pi \right)^8} \frac{1}{2^4} \nonumber \\
      & \times &  \left( dy_1 p_{1t} dp_{1t} d \phi_1 \right) \left( dy_2 p_{2t} dp_{2t} d \phi_2 \right) dy_3 dy_4 d^2p_m  \; .
      \label{mom_space_3}
\end{eqnarray}
The energy-momentum conservation gives the following 
system of equations that has to be solved for discrete solutions
\begin{eqnarray}
\left\{\begin{array}{rll}
&\sqrt{s} - E_3 - E_4& = \sqrt{m_{1t}^2 +p_{1z}^2}+ \sqrt{m_{2t}^2 + p_{2z}^2} \; , \\
& - p_{3z} - p_{4z} &  =  p_{1z} + p_{2z} \;, \\
\end{array}\right.
\label{discrete_solution}
\end{eqnarray}
where $m_{1t}$, $m_{2t}$ are the so-called transverse masses of 
outgoing nuclei which are defined as:
\begin{equation}
m_{it}^2 = p_{it}^2 + m_i^2  \; .
\end{equation}

We wish to make the transformation from 
$\left( y_1, \, y_2 \right)$ to $\left( p_{1z}, \, p_{2z} \right)$.
The transformation Jacobian takes the form:
\begin{equation}
\mathcal{J}_k = \left| \frac{p_{1z} \left(k \right)}{\sqrt{m_{1t}^2 + p_{1z}^2 \left( k \right)}} - \frac{p_{2z} \left(k \right)}{\sqrt{m_{2t}^2 + p_{2z}^2 \left( k \right)}} \right| \; ,
\end{equation}
where $k$ numerates discrete solutions of Eq.~(\ref{discrete_solution}). 
Thus the cross section for the $2 \to 4$ 
reaction reads:
\begin{eqnarray}
\sigma & = & \int \sum_k \mathcal{J}_k^{-1} \left( p_{1t}, \, \phi_1, \, p_{2t}, \, \phi_2, \, y_3, \, y_4, \, p_m, \, \phi_m \right) \frac{1}{2\sqrt{s \left( s -4m^2 \right)}} \overline{ | \mathcal{M} | ^2}  \frac{1}{\left( 2 \pi \right)^8} \frac{1}{2^4} \nonumber \\
      & \times &  \left( p_{1t} dp_{1t} d \phi_1 \right) \left(p_{2t} dp_{2t} d \phi_2 \right) \frac{1}{4} dy_3 dy_4 d^2p_m  \; .
      \label{mom_space_end}
\end{eqnarray}
For photon-exchanges, considered here, it is convenient to change the variables $p_{1t} \to \xi_1 = \log_{10}\left(p_{1t}\right)$, $p_{2t} \to \xi_2 = \log_{10}\left(p_{2t}\right)$.
The lepton helicity dependent amplitudes of the process shown in Fig.~\ref{fig:momentum_space} can be written as:
\begin{eqnarray}
 {\cal M}_{\lambda_3,\lambda_4} \left( t \mbox{-channel} \right) & = & e\, F_{ch} \left( q_1 \right) \, \left( p_a+p_1 \right)^\alpha
 \frac{-i\,g_{\alpha \mu}}{q_1^2+i \varepsilon} \bar{u} \left( p_3,\, \lambda_3 \right) \, i \, \gamma^\mu
 \frac{i\, \left[ \left( \not{p}_3 - \not{q}_1 \right) + m_{\mu} \right]}{\left( q_1-p_3 \right)^2 -m_{\mu}^2} \nonumber \\
 & \times & i \, \gamma^\nu \,  v \left( p_4,\, \lambda_4 \right) \, \frac{-i\,g_{\nu \beta}}{q_2^2+i \varepsilon} \, \left( p_b+p_2 \right)^\beta \, e\, F_{ch} \left( q_2 \right) 
 \label{t_channel}
\end{eqnarray}
and
\begin{eqnarray}
 {\cal M}_{\lambda_3,\lambda_4} \left( u \mbox{-channel} \right) & = & e\, F_{ch} \left( q_1 \right) \, \left( p_a+p_1 \right)^\alpha
 \frac{-i\,g_{\alpha \mu}}{q_1^2+i \varepsilon} \bar{u} \left( p_3,\, \lambda_3 \right) \, i \, \gamma^\nu
 \frac{i\, \left[ \left( \not{p}_3 - \not{q}_2 \right) + m_{\mu} \right]}{\left( q_2-p_3 \right)^2 -m_{\mu}^2} \nonumber \\
 & \times & i \, \gamma^\mu \,  v \left( p_4,\, \lambda_4 \right) \, \frac{-i\,g_{\nu \beta}}{q_2^2+i \varepsilon} \, \left( p_b+p_2 \right)^\beta \, e\, F_{ch} \left( q_2 \right) \; .
 \label{u_channel}
\end{eqnarray}

These amplitudes are calculated numerically.
Finally, to calulate the total cross section one has to calculate 
the 8-dimensional integral inserting 
${\cal M}_{\lambda_3,\lambda_4}={\cal M}_{\lambda_3,\lambda_4} \left( t \mbox{-channel} \right) + {\cal M}_{\lambda_3,\lambda_4} \left( u \mbox{-channel} \right)$ into Eq.~(\ref{mom_space_end}).
We shall compare the impact parameter EPA results 
with the exact 
\footnote{By exact we mean the correct inclusion of the $2 \to 4$ process phase-space. It is, however, rather difficult to include absorption effects in this approach.} Quantum Electrodynamics results.

\section{Results}
\begin{figure}[!h]    
\begin{minipage}[t]{0.46\textwidth}
\centering
\includegraphics[width=1\textwidth]{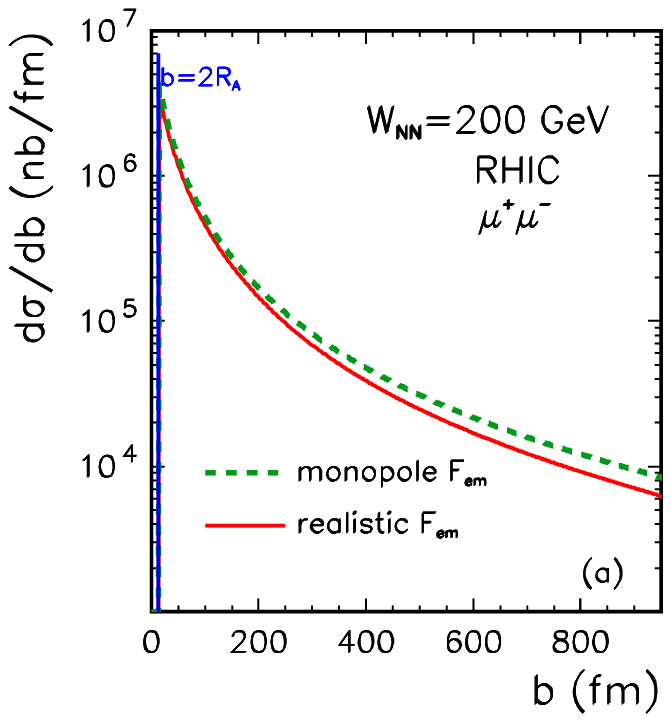}
\end{minipage}
\hspace{0.03\textwidth}
\begin{minipage}[t]{0.46\textwidth}
\centering
\includegraphics[width=1\textwidth]{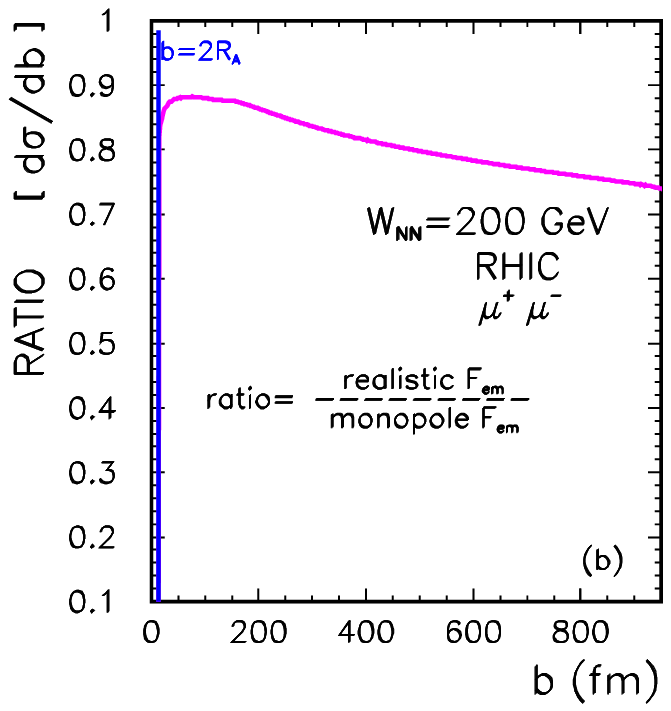}
\end{minipage}
   \caption{\label{fig:dsig_dbm_mu_EPA}
   \small (Color online) The cross section as a function of the impact parameter
for the $AuAu \to \mu^+ \mu^- AuAu$ reaction calculated in 
the equivalent photon approximation.
In the left panel we show the results for realistic charge 
distribution (solid line) and for monopole form factor 
(dashed line). 
On the right side we depict the ratio : 
$RATIO = {\rm d} \sigma \left( F_{em}^{REALISTIC} \right) 
/ {\rm d} \sigma \left( F_{em}^{MONOPOLE} \right)$.
}
\end{figure}

%
Let us start from the presentation of the results obtained in the impact
parameter EPA. In Fig.\ref{fig:dsig_dbm_mu_EPA} we show 
the distribution in the impact parameter $b$ for typical
RHIC energy $\sqrt{s_{NN}}$ = 200 GeV. 
The contributions from distances smaller than $b=2R_A$ are cut off and 
consistently with $\theta$-function in Eq.~(\ref{eq.tot_cross_section_our}).
We clearly see a huge
contribution from distances large compared to the nuclear size.
The distribution with realistic charge falls off somewhat quicker
than that for the monopole charge form factor. This is better 
visualized in the right panel where the ratio of the corresponding 
cross sections is shown. 
\begin{figure}[!h]    
\begin{minipage}[t]{0.46\textwidth}
\centering
\includegraphics[width=1\textwidth]{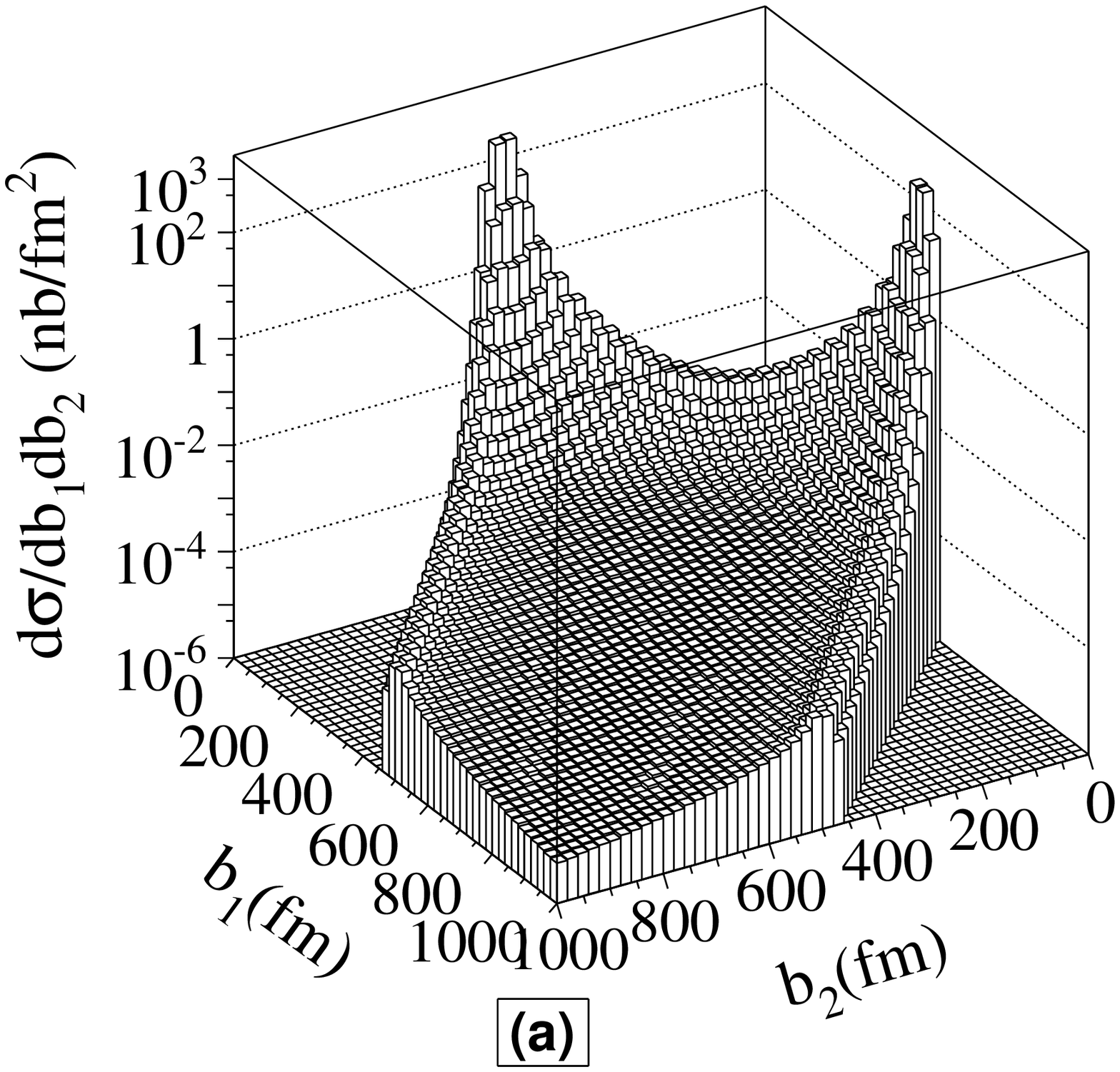}
\end{minipage}
\hspace{0.03\textwidth}
\begin{minipage}[t]{0.46\textwidth}
\centering
\includegraphics[width=1\textwidth]{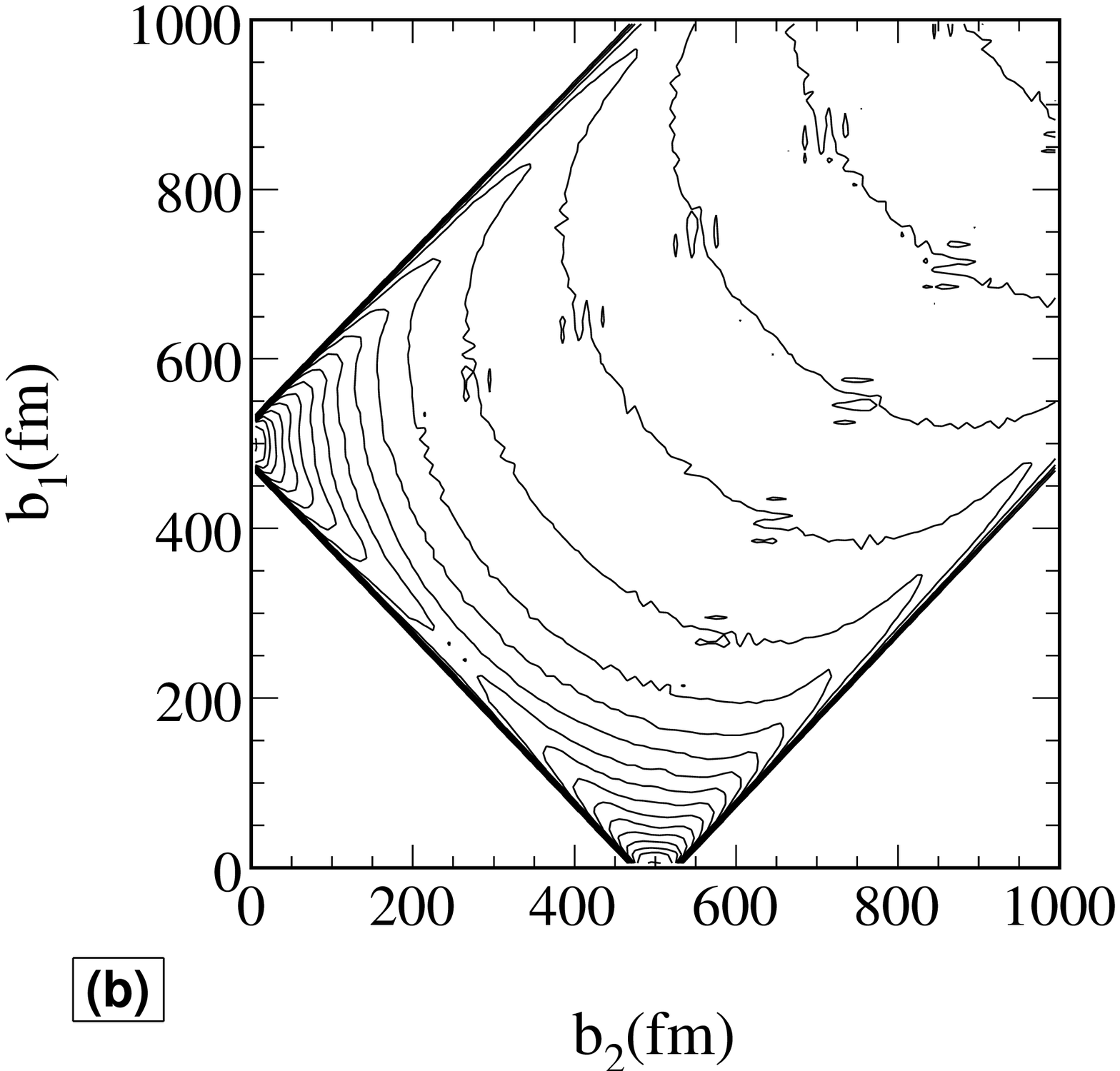}
\end{minipage}
   \caption{\label{fig:production_region}
   \small $\frac{{\rm d} \sigma}{{\rm d} b_1 {\rm d} b_2}$ as a function of $b_1$ and $b_2$ in lego (left) and contour (right) representation for $b \in$ (480,520) fm.
}
\end{figure}

The difference of the cross sections
for the monopole and exact charge form factors
at large impact parameter $b$ shown in the figure is 
especially intriguing in the light of the equality of 
the photon flux factors at large $b_1$ or $b_2$ 
(see Fig.\ref{fig:dflux}).
How to understand this quite nonintuitive result?
In Fig.\ref{fig:production_region} we show 
the distribution of $d\sigma/db_1 db_2$
in ($b_1, b_2$) with the severe restriction for the impact parameter
 $b \in$ (480,520) fm. We see two pronounced peaks
at ($b_1 \approx b,b_2 \approx 0$) and
($b_1 \approx 0, b_2 \approx b$).
This demonstrates a strong preference of asymmetric 
production of the pair: close to the trajectory of one 
or the other nucleus, where the form factor
details are important (see Fig.\ref{fig:dflux}).
This point was never discussed so far in 
the literature.

\begin{figure}[!h]    
\begin{minipage}[t]{0.46\textwidth}
\centering
\includegraphics[width=1\textwidth]{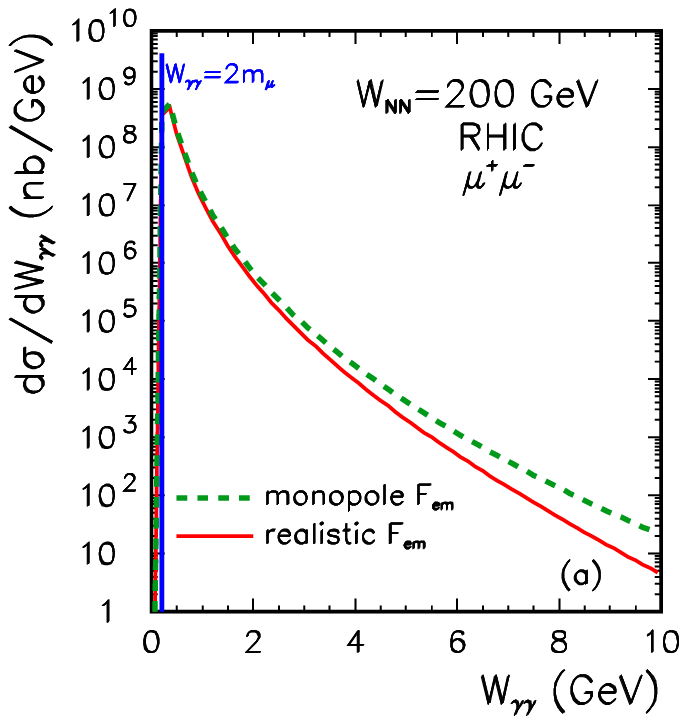}
\end{minipage}
\hspace{0.03\textwidth}
\begin{minipage}[t]{0.46\textwidth}
\centering
\includegraphics[width=1\textwidth]{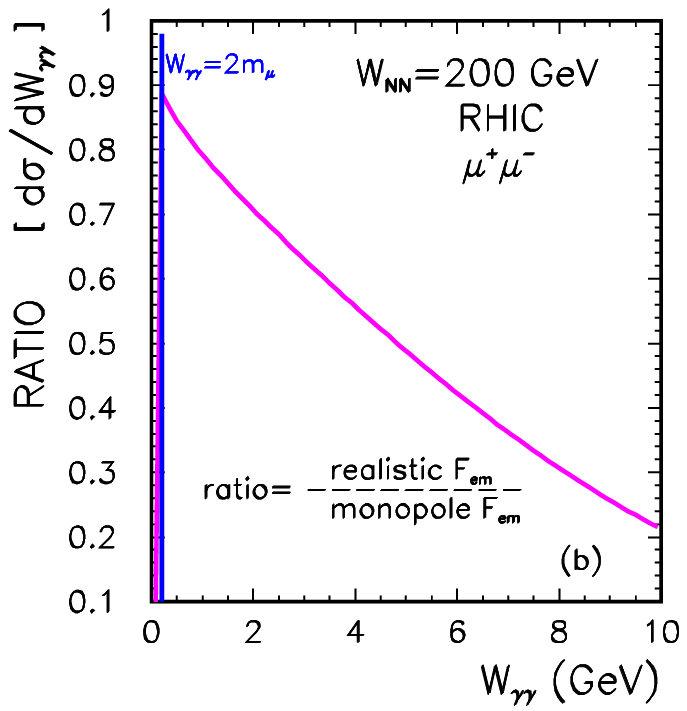}
\end{minipage}
   \caption{\label{fig:dsig_dw_mu_EPA}
   \small (Color online) The cross section for $Au- Au$ scattering 
as a function of photon--photon center--of--mass energy 
$W_{\gamma \gamma}=M_{\mu^+ \mu^-}$ in EPA. In the right panel we show 
the ratio of "realistic" to "monopole" form factor.
}
\end{figure}

The distributions shown in Fig.\ref{fig:dsig_dbm_mu_EPA} 
are purely theoretical, that is cannot be easily measured. 
Let us come now to the distributions which could, at least 
in principle, be measured.
Fig.\ref{fig:dsig_dw_mu_EPA} shows the distribution in the dimuon
subsystem energy. The distributions in $W_{\gamma \gamma}=M_{\mu^+ \mu^-}$ 
falls steeply off.
In the right panel we show the ratio of the cross
sections for realistic charge distribution to that for the monopole
charge form factor. At $W_{\gamma \gamma}$ = 10 GeV the two 
distributions differ already by a factor of about 5 which clearly 
shows limitations of the calculations with analytic charge form factors.

\begin{figure}[!h]    
\begin{minipage}[t]{0.46\textwidth}
\centering
\includegraphics[width=1\textwidth]{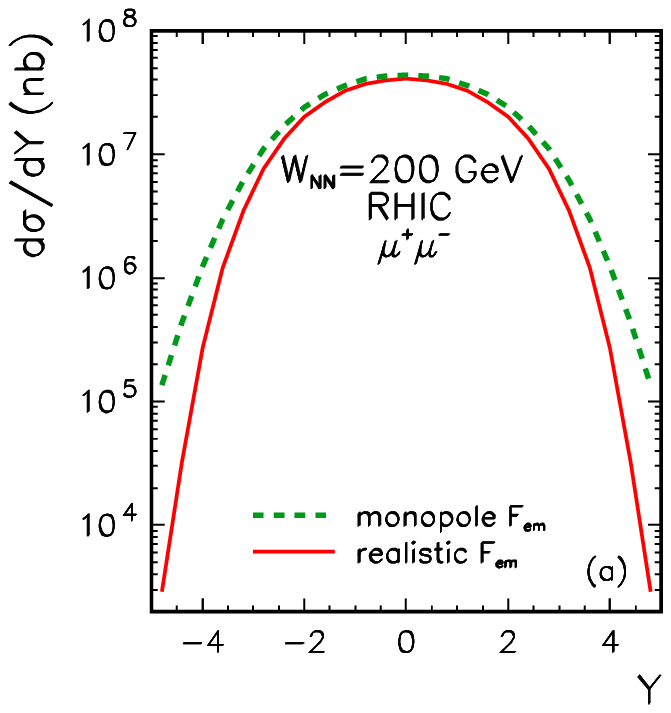}
\end{minipage}
\hspace{0.03\textwidth}
\begin{minipage}[t]{0.46\textwidth}
\centering
\includegraphics[width=1\textwidth]{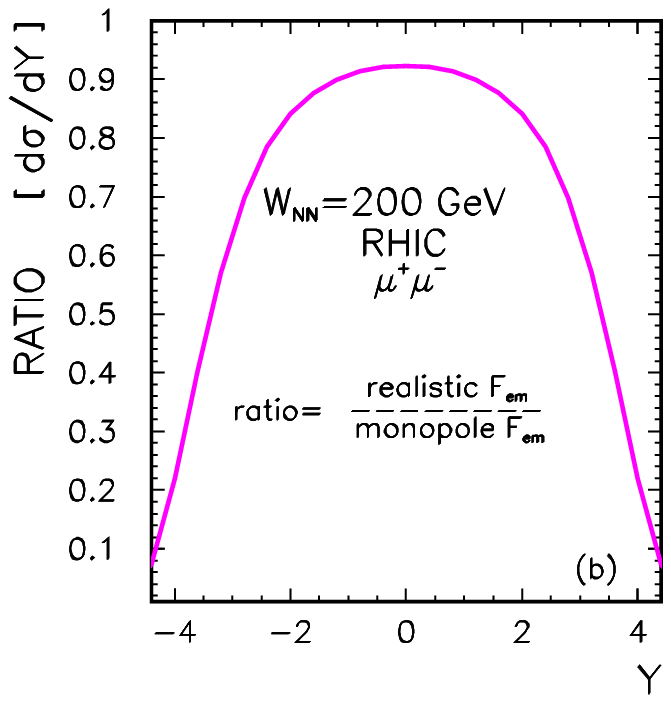}
\end{minipage}
   \caption{\label{fig:dsig_dy_mu_EPA}
   \small (Color online) The cross section as a function of 
$Y=\frac{1}{2} \left( y_{\mu^+} + y_{\mu^-} \right)$ (left panel)
for realistic and monopole form factors (left)
calculated in EPA and their ratio (right).
}
\end{figure}

Finally, in analogy to the $A A \to A A \rho^0 \rho^0$ reaction
studied in Ref.\cite{KSS09}, in Fig.\ref{fig:dsig_dy_mu_EPA}
we show the distribution in the dimuon pair rapidity. 
As for the $\rho^0 \rho^0$ production we see a huge difference between 
the results of the two calculations for large dilepton rapidities.
Measurements of dileptons in forward directions would be therefore very useful
to understand the role of realistic charge distribution. The relative effect
is shown in the right panel of the figure.

The preliminary calculation in the impact parameter space clearly
shows how important can be studying of differential distributions 
to pin down the effects of realistic charge density.
Not all of the distributions can be easily addressed in the impact parameter 
approach.
The Feynman diagram approach in the momentum space seems to be a better
alternative to study the differential distributions.

\begin{figure}[!h]    
\begin{minipage}[t]{0.46\textwidth}
\centering
\includegraphics[width=1\textwidth]{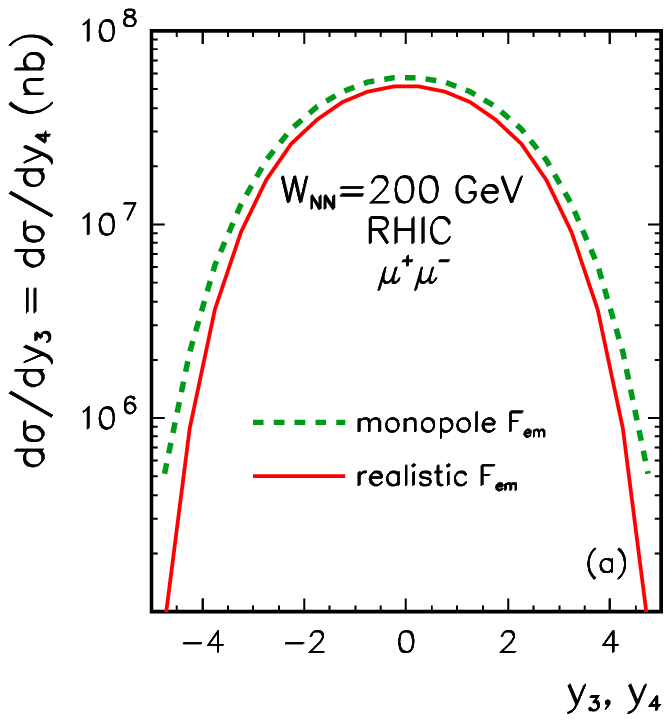}
\end{minipage}
\hspace{0.03\textwidth}
\begin{minipage}[t]{0.46\textwidth}
\centering
\includegraphics[width=1\textwidth]{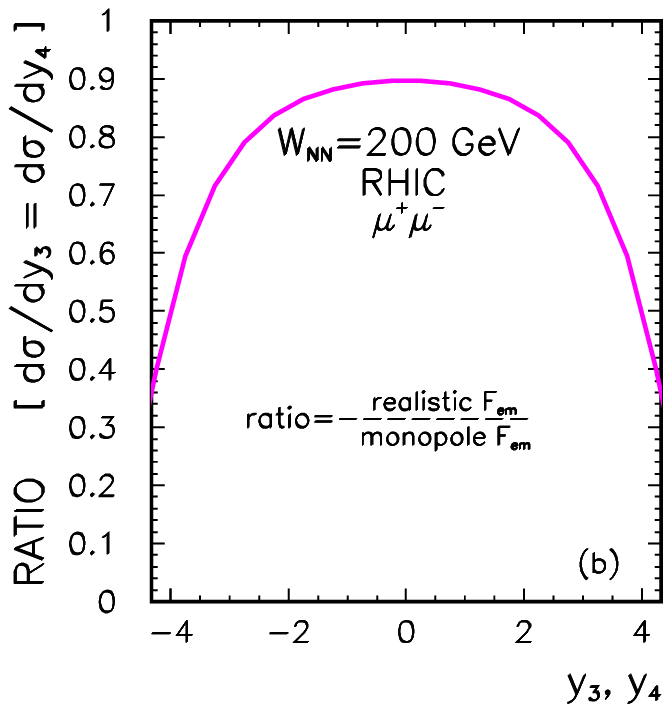}
\end{minipage}
   \caption{\label{fig:dsig_dy3_ms}
   \small (Color online) The cross section as a function of $y_{\mu^+}$, $y_{\mu^-}$
for realistic and monopole form factor calculated in the momentum
space (left panel). Their ratio is shown in the right panel.
}
\end{figure}

Now we come to the presentation of results obtained in the momentum
space approach with details outlined in Section II.
Fig.\ref{fig:dsig_dy3_ms} shows distributions in muon rapidities
(identical for $\mu^+$ and $\mu^-$). No other limitations or kinematical cuts have been
included here. As in the previous cases we show distributions 
obtained with the monopole and realistic charge form factor. 
The effect of the oscillatory character of $F_{ch}(q)$
and in particular its first minimum is reflected by a smaller 
cross section at larger rapidities compared 
to the results obtained with monopole form factor. This is due to the fact 
that on average at large rapidities larger four-momentum squared transfers 
($t_1$ or $t_2$) are involved. In reality, one effectively 
integrates over a certain range of $t_1$ and $t_2$. 
The relative effect is shown in the right panel.

\begin{figure}[!h]    
\begin{minipage}[t]{0.46\textwidth}
\centering
\includegraphics[width=1\textwidth]{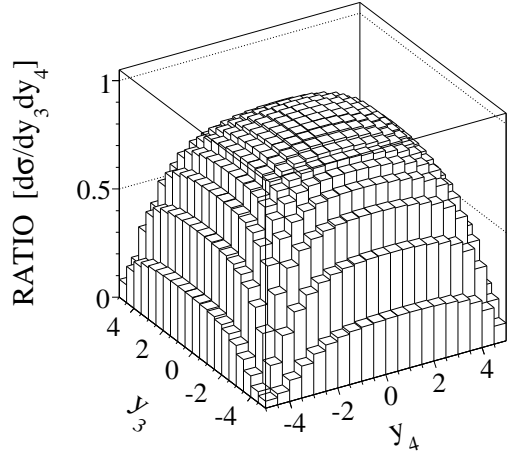}
\end{minipage}
   \caption{\label{fig:ratio_y3y4_ms}
   \small The ratio of two-dimensial distributions 
   ${\rm d} \sigma \left( F_{em}^{REALISTIC} \right) 
/ {\rm d} \sigma \left( F_{em}^{MONOPOLE} \right)$ in
$y_3$ and $y_4$.
}
\end{figure}

Fig.\ref{fig:ratio_y3y4_ms} shows the situation (the ratio
of the two calculations) in the two dimensional space: 
($y_3,y_4$).
Clearly at mid rapidities, where on average rather small $t_1$
and $t_2$ are involved, the use of the approximate monopole form 
factor is justified. This is not the case at the edges of the 
($y_3,y_4$) plane where due to kinematics $\left|t_1 \right|$ or/and $\left|t_2 \right|$
are larger.

Up to now we have discussed "a theoretical situation" when all 
the muons are accepted. In practice one can measure only muons
with transverse momenta larger than a certain value,
characteristic for a given detector.
We shall consider now cases relevant for concrete experimental
situations.

\begin{figure}[!h]    
\begin{minipage}[t]{0.46\textwidth}
\centering
\includegraphics[width=1\textwidth]{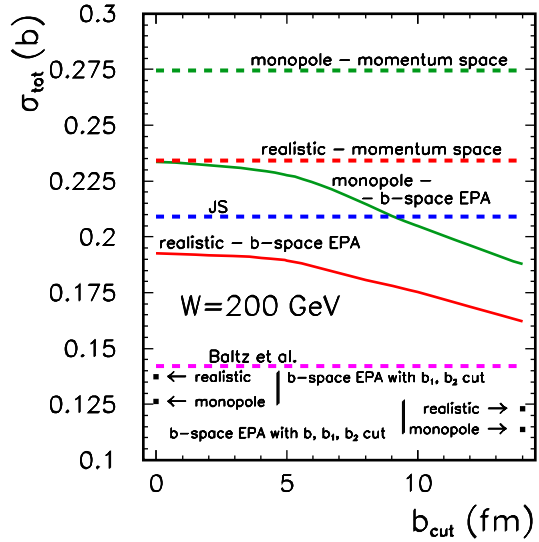}
\end{minipage}
   \caption{\label{fig:sigma_tot_b}
   \small (Color online) The compilation of the results obtained in different approaches 
   for the total cross section for 
    $Au Au \to Au Au \mu^+ \mu^-$ at $\sqrt{s}_{NN}$ = 200 GeV.
}
\end{figure}

The calculations in the literature concentrated mostly on the total cross section.
In Fig.\ref{fig:sigma_tot_b} we present the dependence of the total cross section
on the lower cut-off in the impact parameter. We present EPA results for realistic
(lower solid line) and monopole (upper solid line) form factors.
The cross section without the cut-off is by 15\% larger than that for $b_{cut}$ = 14 fm.
This result is smaller than the corresponding results obtained
within momentum space calculations, shown as the horizontal dashed lines.
Different methods has been used in the literature to calculate the total
cross section for the $AuAu \to AuAu\mu^+ \mu^-$ process. 
For comparison we show also results obtained recently by Jentschura and Serbo (JS) 
\cite{SB} in the momentum space EPA and by Baltz et al. \cite{BGKN09} in 
the b-space EPA. The JS result should be compared to our momentum space
calculation with monopole form factor. Our exact calculation is in this case
larger than their EPA calculation by about 24\%. This shows the precision
of the momentum space EPA.
The Baltz et al. result is significantly lower than our b-space EPA result.
In their calculations the cuts were imposed rather on $b_1$ and $b_2$, instead
on $b$ in our case. If we impose additional cuts on $b_1$ and $b_2$ in 
Eq.~(\ref{eq.tot_cross_section_our}) we get the point in the lower-right corner. If the cut on $b$ is
not imposed we get the point in the lower-left corner. The result of Baltz et al.
differs from both these values, the solution being most probably a different
form factor used in their case.

\begin{figure}[!h]    
\begin{minipage}[t]{0.46\textwidth}
\centering
\includegraphics[width=1\textwidth]{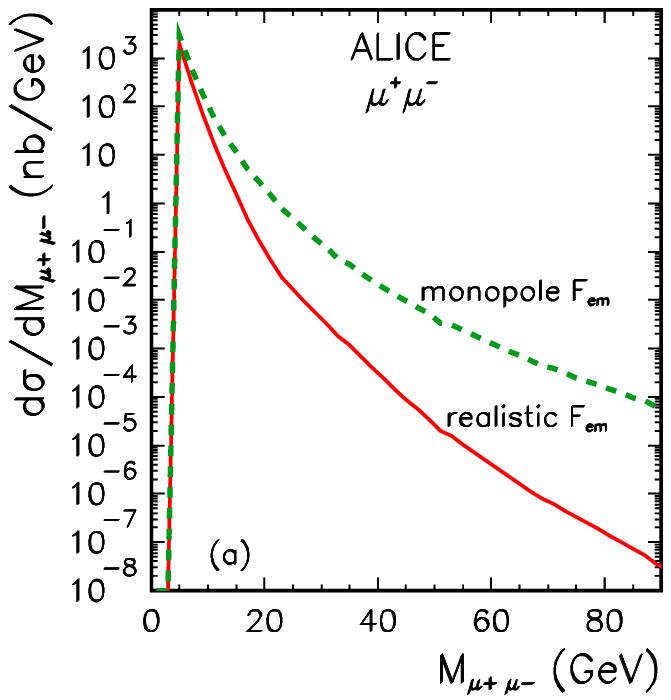}
\end{minipage}
\hspace{0.03\textwidth}
\begin{minipage}[t]{0.46\textwidth}
\centering
\includegraphics[width=1\textwidth]{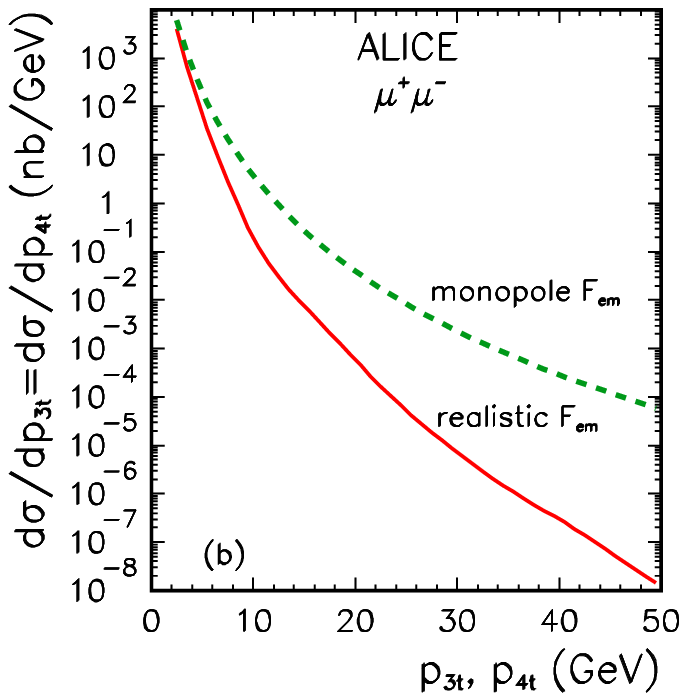}
\end{minipage}
   \caption{\label{fig:ALICE_varia}
   \small 
(Color online) Invariant mass distribution 
$\frac{{\rm d} \sigma}{{\rm d} M_{\mu^+ \mu^-}}$ (left) and
muon transverse momentum distribution 
$\frac{{\rm d} \sigma}{{\rm d} p_{3t}} = \frac{{\rm d} \sigma}{{\rm d} p_{4t}}$ (right)
for ALICE conditions: $y_3, y_4 \in (3,4)$, $p_{3t}, p_{4t} \geq$ 2 GeV 
and the center-of-mass energy $W_{NN}=5.5$ TeV.
}
\end{figure}

Now we will continue reviewing our predictions for the differential distributions.
Let us start with the ALICE detector. The ALICE collaboration can
measure only forward muons with psudorapidity 4 $< \eta <$ 5
and uses a relatively low cut on muon transverse momentum, 
$p_t >$ 2 GeV.
In Fig.\ref{fig:ALICE_varia} (left panel) we show the invariant mass 
distribution of dimuons for monopole and realistic form factors. 
The ALICE experimental cuts were incorporated into our calculations. 
The bigger invariant mass the bigger the difference
between the results for the two form factors.
The same is true for distributions in muon transverse
momenta (see the right panel). 

\begin{figure}[!h]    
\begin{minipage}[t]{0.46\textwidth}
\centering
\includegraphics[width=1\textwidth]{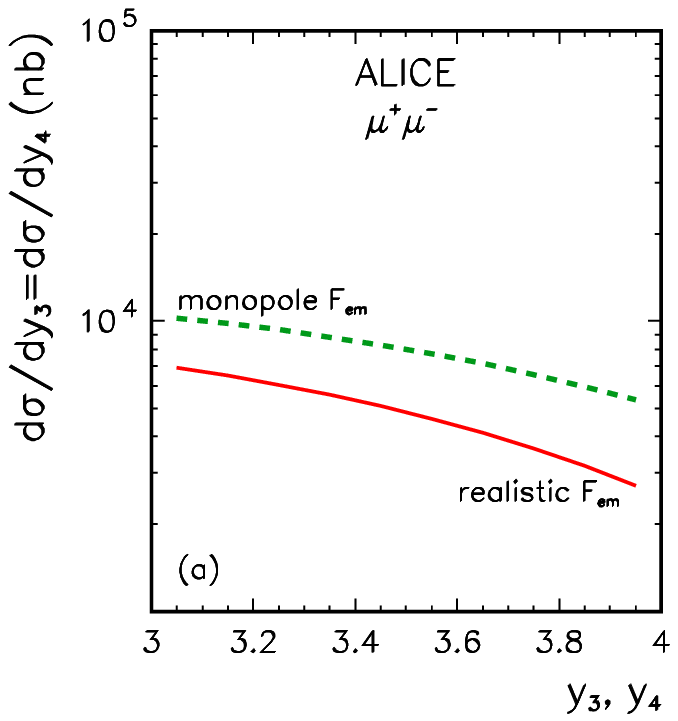}
\end{minipage}
\hspace{0.03\textwidth}
\begin{minipage}[t]{0.46\textwidth}
\centering
\includegraphics[width=1\textwidth]{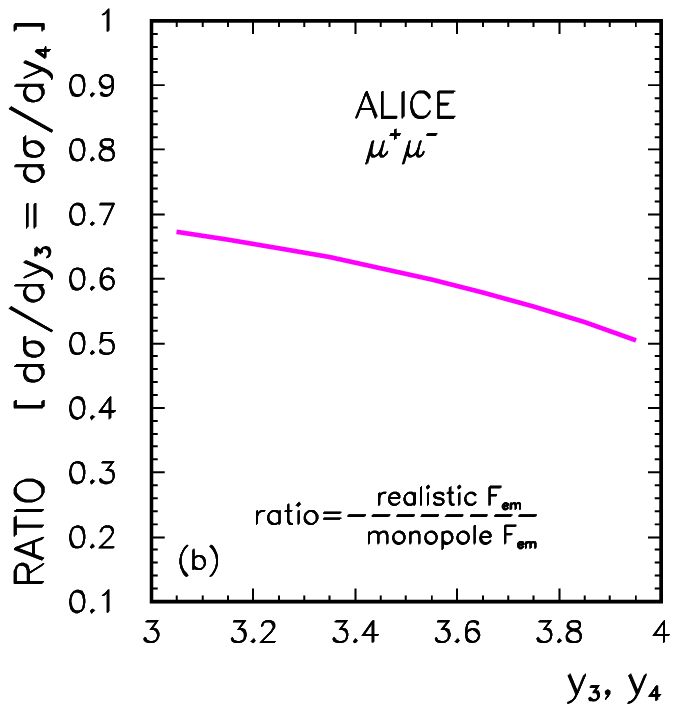}
\end{minipage}
   \caption{\label{fig:dsig_dy3_ALICE}
   \small 
(Color online) $\frac{{\rm d} \sigma}{{\rm d} y_3} = \frac{{\rm d} \sigma}{{\rm d} y_4}$
(left) and their ratio (right) for the ALICE conditions:  
$y_3, y_4 \in (3,4)$, $p_{3t}, p_{4t} \geq$ 2 GeV and $W_{NN}=5.5$ TeV.}
\end{figure}

The distribution in rapidity is shown in 
Fig.\ref{fig:dsig_dy3_ALICE}. We present the cross sections for 
both (realistic, monopole) form factors and their ratio.

\begin{figure}[!h]    
\begin{minipage}[t]{0.46\textwidth}
\centering
\includegraphics[width=1\textwidth]{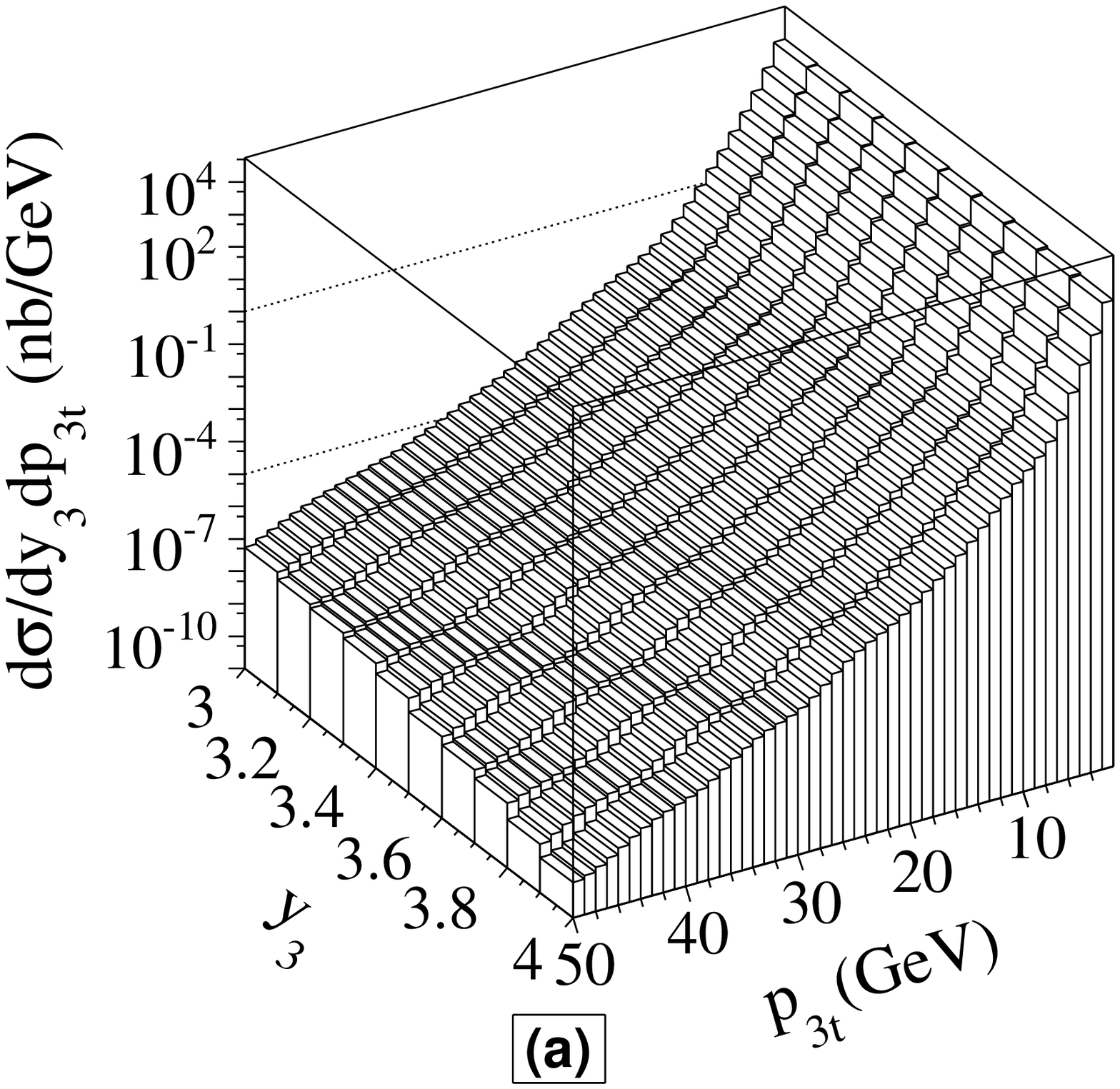}
\end{minipage}
\hspace{0.03\textwidth}
\begin{minipage}[t]{0.46\textwidth}
\centering
\includegraphics[width=1\textwidth]{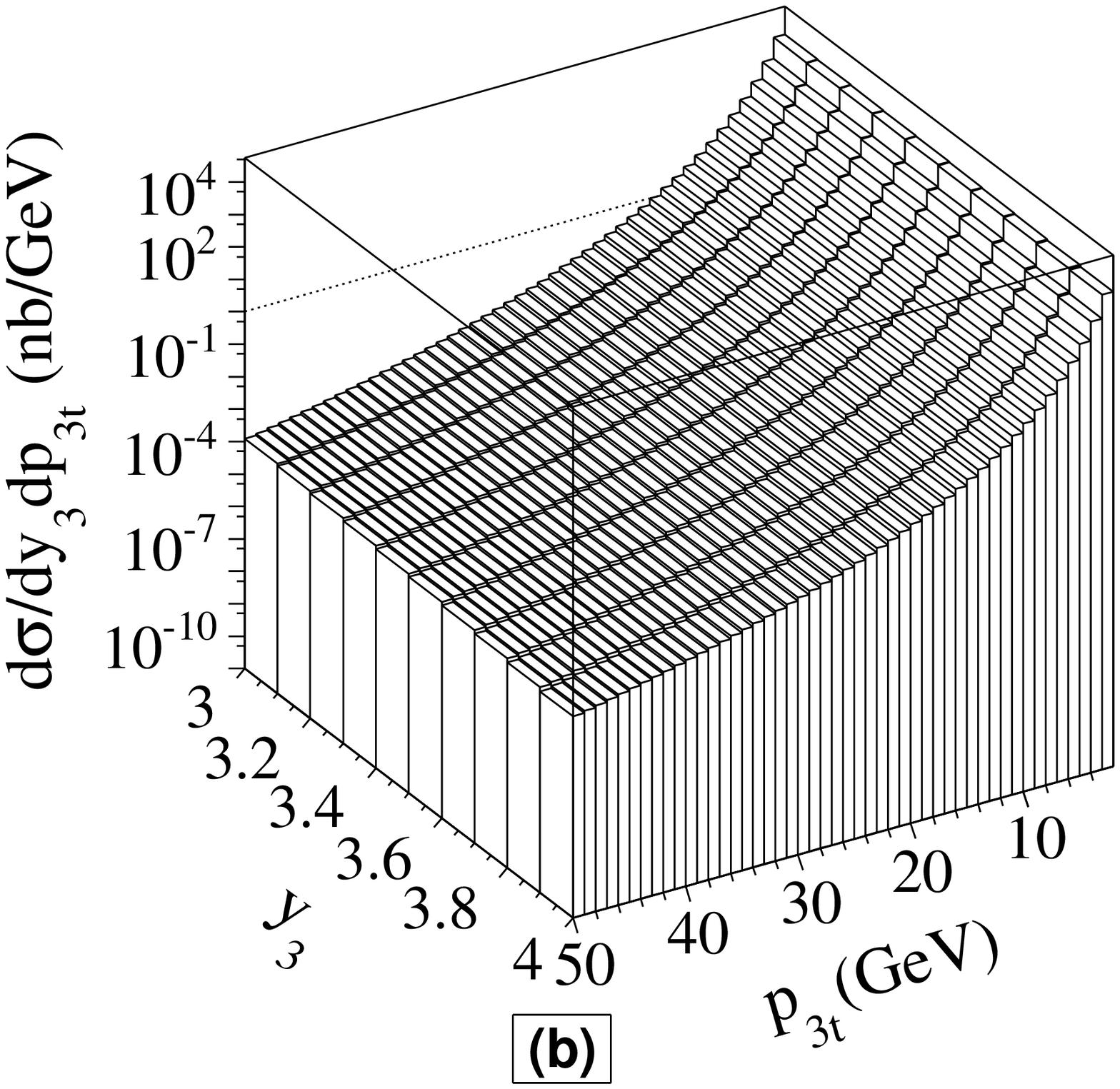}
\end{minipage}
   \caption{\label{fig:y3p3t_ALICE}
   \small 
Double differential cross section $\frac{{\rm d} \sigma}{{\rm d} y_3 \, {\rm d} p_{3t}}$ for realistic (left)
and monopole (right) form factors for 
ALICE conditions $y_3, y_4 \in (3,4)$, $p_{3t}, p_{4t} \geq$ 2 GeV and $W_{NN}=5.5$ TeV.}
\end{figure}

Double differential distribution of the muon rapidity and
transverse momentum is shown in 
Fig.\ref{fig:y3p3t_ALICE}. These are our predictions
which could be studied experimentally in the future. 
The small irregularities seen in the two-dimensional spectra 
for realistic form factor are the consequence of 
the oscillatory character of the nucleus charge form factor.
The distribution for the monopole form factor is more smooth.

\begin{figure}[!h]    
\begin{minipage}[t]{0.46\textwidth}
\centering
\includegraphics[width=1\textwidth]{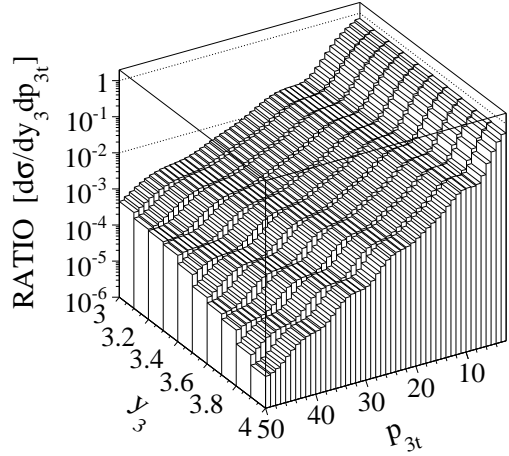}
\end{minipage}
   \caption{\label{fig:ratio_y3p3t_ALICE}
   \small 
Ratio of the cross sections 
$\frac{{\rm d} \sigma}{{\rm d} y_3 \, {\rm d} p_{3t}}$ 
for the ALICE conditions: $y_3, y_4 \in (3,4)$, $p_{3t}, p_{4t} \geq$
 2 GeV and $W_{NN}=5.5$ TeV.
}
\end{figure}

In Fig.\ref{fig:ratio_y3p3t_ALICE} we show the ratio of 
the cross sections shown in the previous figure. 
Huge deviations from the unity can be seen. 
The reminiscence of the oscillating form factor can be seen also 
in the ratio. Experimental confirmation of this behaviour 
would be very useful. Moreover it would demonstrate whether
our understanding of the nuclear effects is correct.
Large deviations from the predictions presented here would be 
surprising.

\begin{figure}[!h]    
\begin{minipage}[t]{0.46\textwidth}
\centering
\includegraphics[width=1\textwidth]{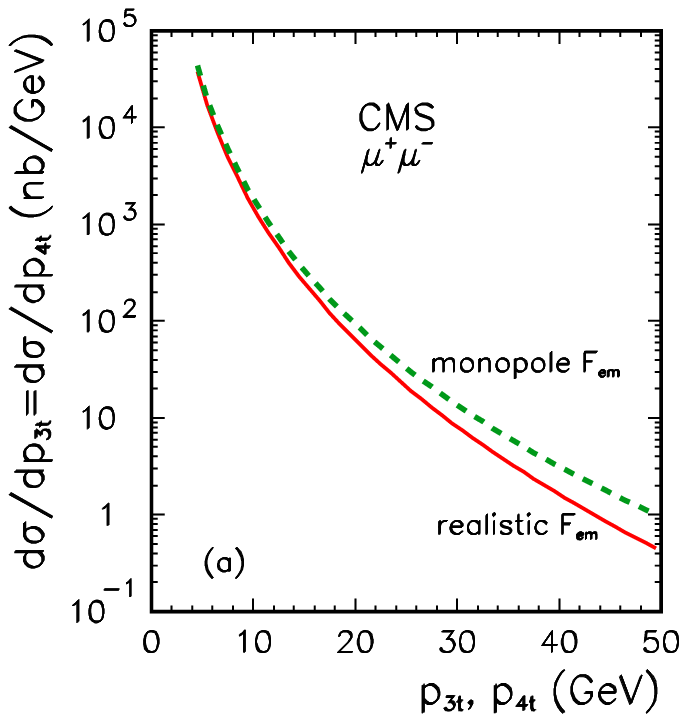}
\end{minipage}
\hspace{0.03\textwidth}
\begin{minipage}[t]{0.46\textwidth}
\centering
\includegraphics[width=1\textwidth]{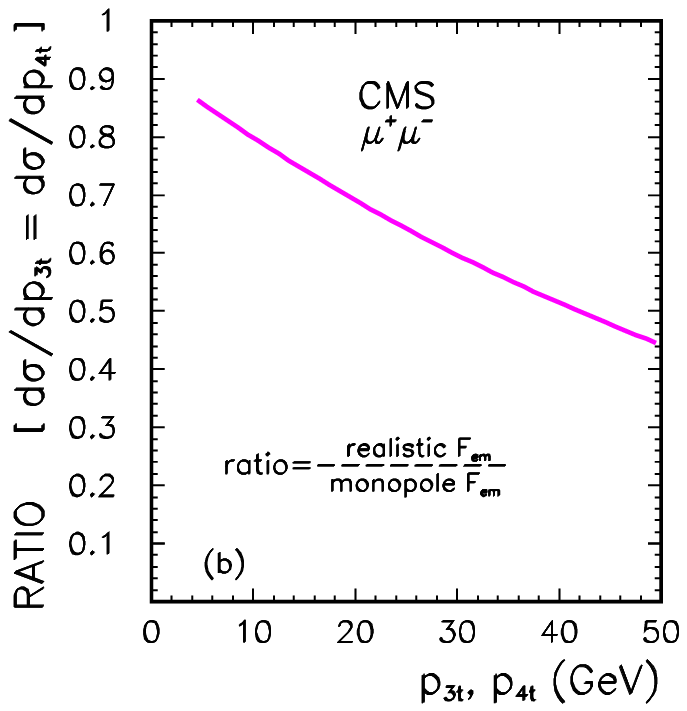}
\end{minipage}
   \caption{\label{fig:dsig_dp3t_CMS}
   \small 
(Color online) The muon transverse momentum distribution 
$\frac{{\rm d} \sigma}{{\rm d} p_{3t}}$ (left) and the ratio (right)
for the CMS conditions: 
$y_3, y_4 \in (-2.5,2.5)$, $p_{3t}, p_{4t} \geq 4$ GeV and $W_{NN}=5.5$ TeV.
}
\end{figure}

Let us come now to the predictions for the CMS detector.
In contrast to the ALICE detector, CMS can measure midrapidity
values with  -2.5 $< y_3, y_4 <$ 2.5. At midrapidities one
samples on average smaller $t_1$ and $t_2$ therefore
the efects of the realistic form factors are expected to be smaller.
Fig.\ref{fig:dsig_dp3t_CMS} confirms the expectations.
Even for muon transverse momenta of 50 GeV one obtains
damping with respect to the result obtained with the monopole
form factor by a factor of about two only.

\begin{figure}[!h]    
\begin{minipage}[t]{0.46\textwidth}
\centering
\includegraphics[width=1\textwidth]{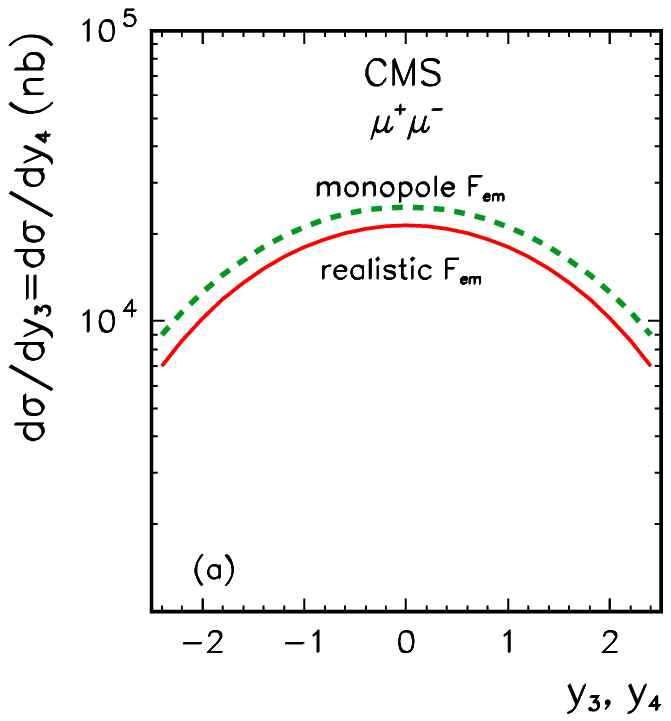}
\end{minipage}
\hspace{0.03\textwidth}
\begin{minipage}[t]{0.46\textwidth}
\centering
\includegraphics[width=1\textwidth]{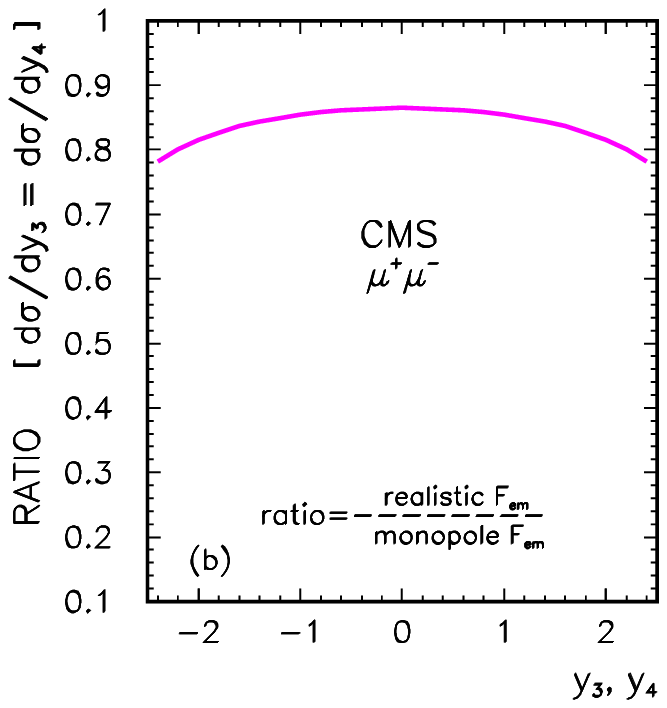}
\end{minipage}
   \caption{\label{fig:dsig_dy3_CMS}
   \small 
(Color online) The muon rapidity distribution $\frac{{\rm d} \sigma}{{\rm d} y_3}$ (left) and the ratio (right)
for the CMS conditions: 
$y_3, y_4 \in (-2.5,2.5)$, $p_{3t}, p_{4t} \geq$ 4 GeV and $W_{NN}=5.5$ TeV.
}
\end{figure}

The cross section dependence on the muon rapidity is shown in 
Fig.\ref{fig:dsig_dy3_CMS}. Rather large cross section of
the order of 0.1 mb is expected within the CMS acceptance.
The average deviation with respect to the monopole form factor
is about 20\% (see the left panel). 

\begin{figure}[!h]    
\begin{minipage}[t]{0.46\textwidth}
\centering
\includegraphics[width=1\textwidth]{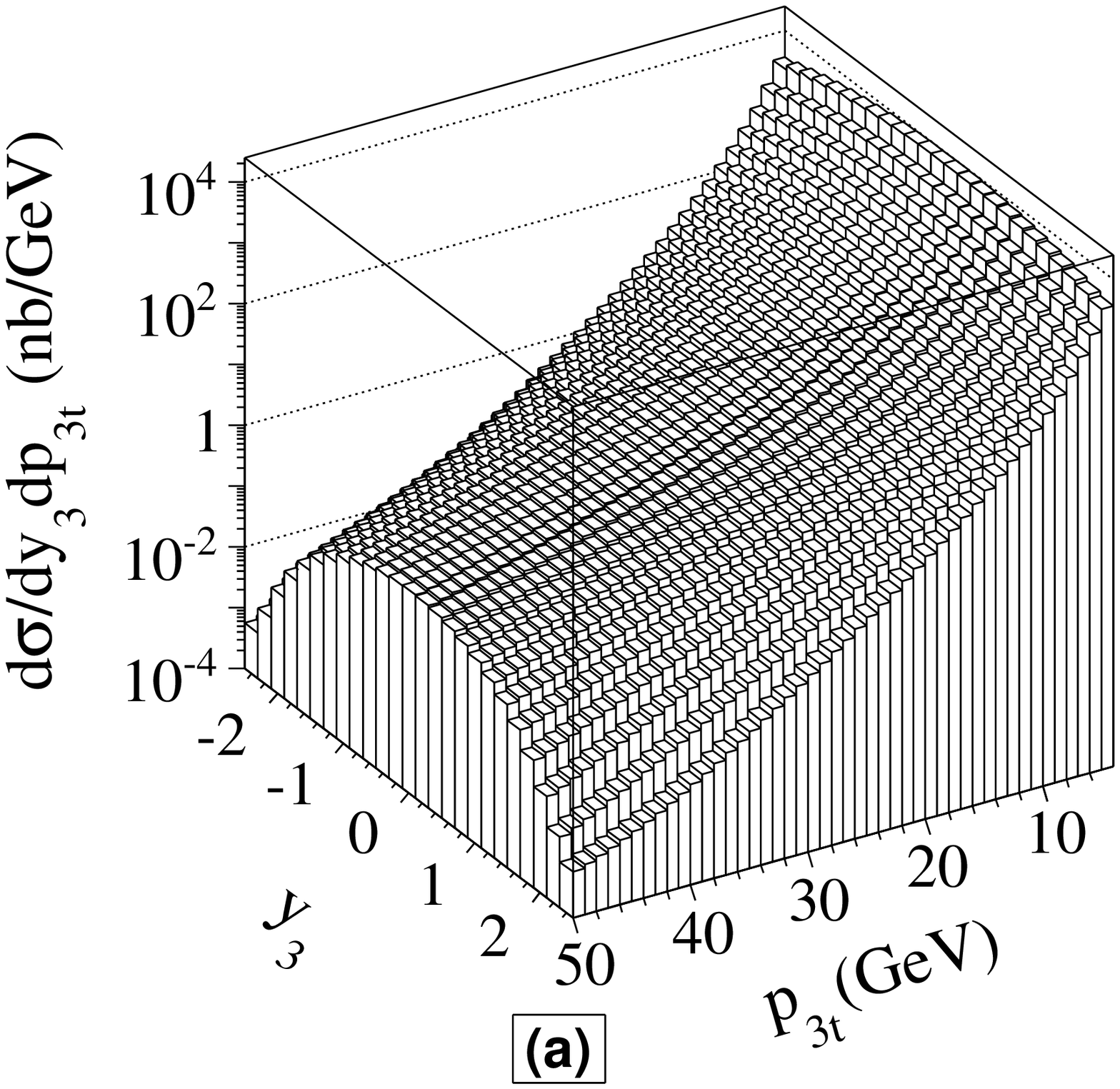}
\end{minipage}
\hspace{0.03\textwidth}
\begin{minipage}[t]{0.46\textwidth}
\centering
\includegraphics[width=1\textwidth]{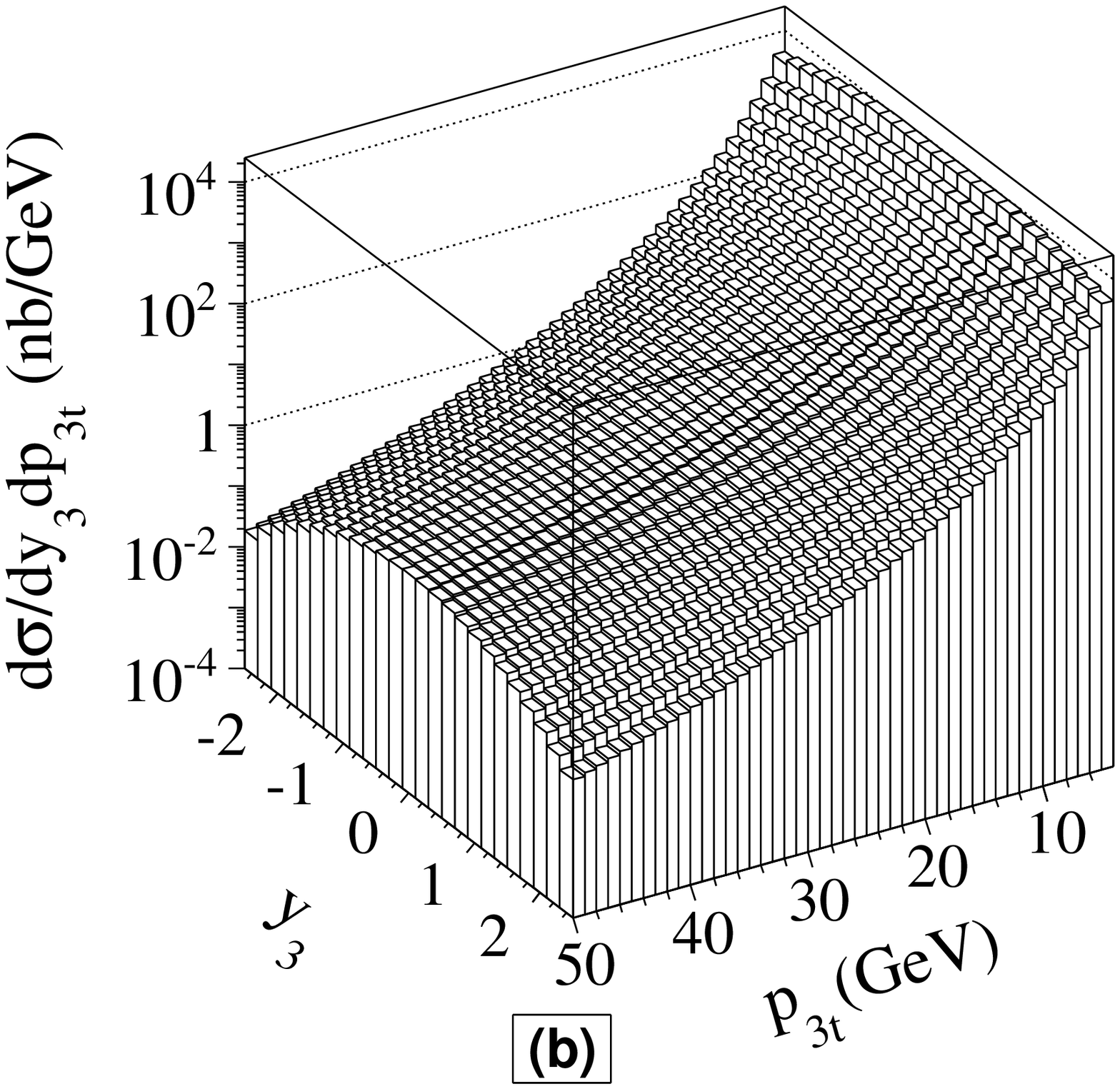}
\end{minipage}
   \caption{\label{fig:y3p3t_CMS}
   \small
$\frac{{\rm d} \sigma}{{\rm d} y_3 \, {\rm d} p_{3t}}$
for realistic (left) and monopole (right) form factors 
for the CMS conditions: 
$y_3, y_4 \in (-2,5.2,5)$, $p_{3t}, p_{4t} \geq$ 4 GeV and $W_{NN}=5.5$ TeV.
}
\end{figure}

\begin{figure}[!h]    
\begin{minipage}[t]{0.46\textwidth}
\centering
\includegraphics[width=1\textwidth]{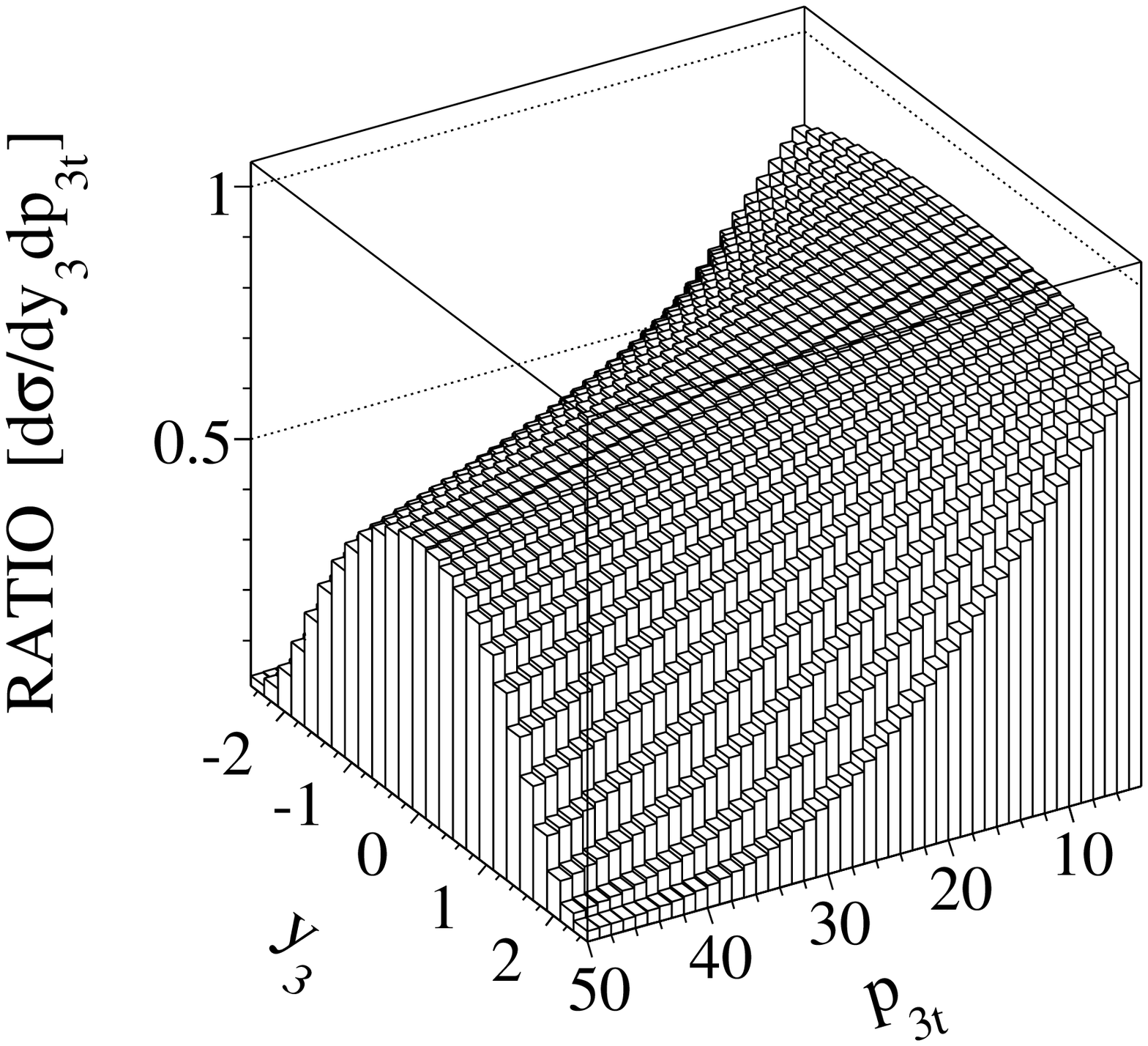}
\end{minipage}
   \caption{\label{fig:dratio_y3p3t_CMS}
   \small 
The ratio of the realistic and monopole cross sections 
$\frac{{\rm d} \sigma}{{\rm d} y_3 \, {\rm d} p_{3t}}$
for the CMS conditions: 
$y_3, y_4 \in (-2.5,2.5)$, $p_{3t}, p_{4t} \geq$ 4 GeV and $W_{NN}=5.5$ TeV.
}
\end{figure}

The two-dimensional distributions within the main CMS detector
are shown in Fig.\ref{fig:y3p3t_CMS}. Big modifications
with respect to the monopole case can be seen for
large $p_t$ and $\left| y_{\mu^+}, y_{\mu^-} \right| \sim$ 2.5, which is also presented
in the form of the ratio in Fig.\ref{fig:dratio_y3p3t_CMS}.


\begin{figure}[!h]    
\begin{minipage}[t]{0.46\textwidth}
\centering
\includegraphics[width=1\textwidth]{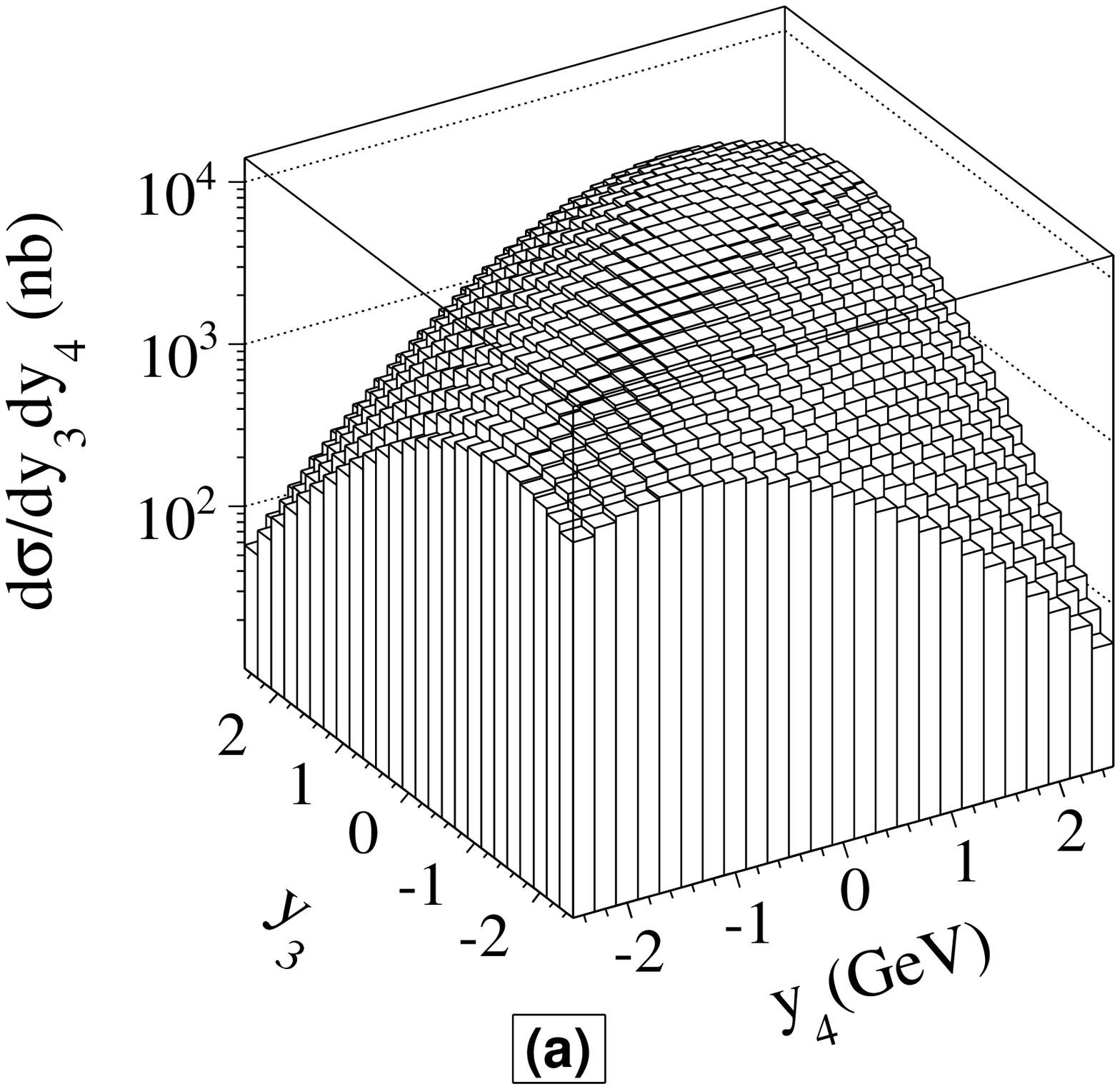}
\end{minipage}
\hspace{0.03\textwidth}
\begin{minipage}[t]{0.46\textwidth}
\centering
\includegraphics[width=1\textwidth]{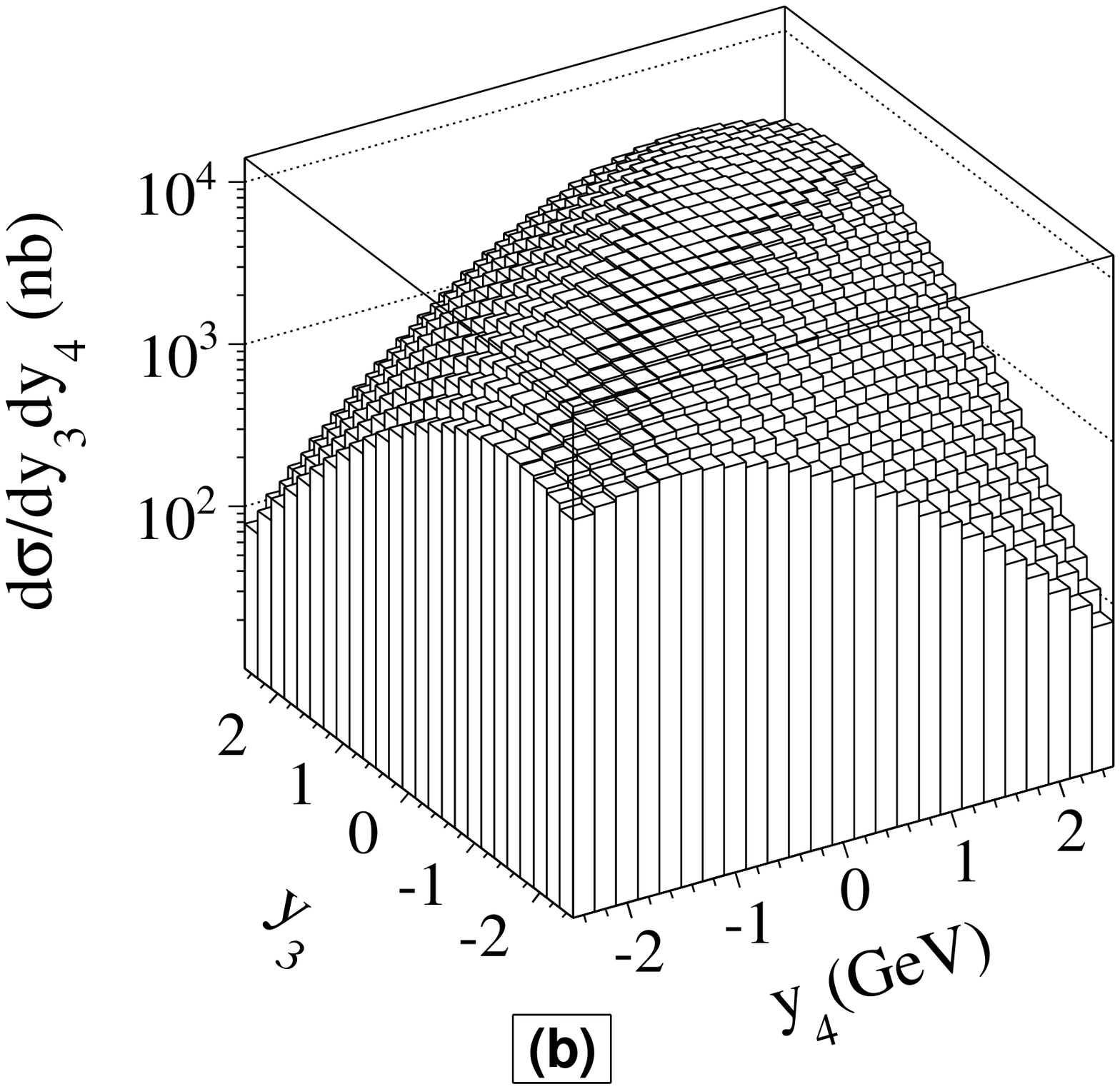}
\end{minipage}
   \caption{\label{fig:y3y4_CMS}
   \small 
$\frac{{\rm d} \sigma}{{\rm d} y_3 \, {\rm d} y_4}$ for realistic (left)
and monopole (right) form factors for
the CMS conditions: 
$y_3, y_4 \in (-2.5,2.5)$, $p_{3t}, p_{4t} \geq$ 4 GeV and $W_{NN}=5.5$ TeV.
}
\end{figure}

Finally, for completeness in Fig.\ref{fig:y3y4_CMS} we show the distributions in 
the $(y_3, y_4)$ plane. Here the distributions
obtained with the monopole and realistic form factors are
rather similar, but one should realize that these distributions
are dominated by muons with small transverse momenta that are 
only slightly bigger than the experimental acceptance $p_t > 2.5$ GeV and as a 
consequence relatively small $t_1$ and $t_2$ values.

\begin{figure}[!h]    
\begin{minipage}[t]{0.46\textwidth}
\centering
\includegraphics[width=1\textwidth]{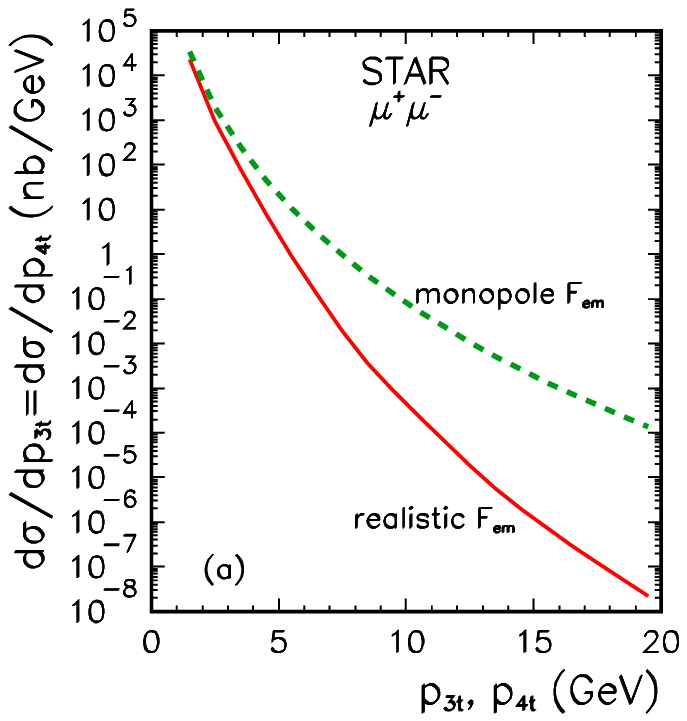}
\end{minipage}
\hspace{0.03\textwidth}
\begin{minipage}[t]{0.46\textwidth}
\centering
\includegraphics[width=1\textwidth]{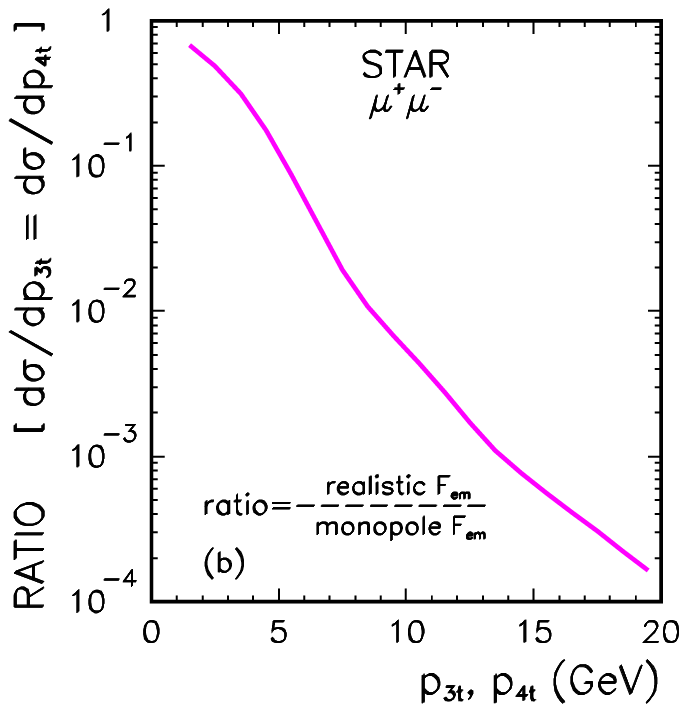}
\end{minipage}
   \caption{\label{fig:dsig_dp3t_STAR}
   \small 
(Color online) $\frac{{\rm d} \sigma}{{\rm d} p_{3t}}$ (left) and the ratio (right)
for the STAR conditions: $y_3, y_4 \in (-1,1)$, $p_{3t}, p_{4t} \geq$
 1 GeV and $W_{NN}=200$ GeV.
}
\end{figure}

The same processes can be also studied at the being presently
in the operation RHIC. Here STAR and PHENIX detectors can be used. 
The distribution of the muon transverse momentum
is shown in Fig.\ref{fig:dsig_dp3t_STAR}. The STAR rapidity cuts
-1 $< y_3, y_4 <$ 1 are taken into account. Compared
to the LHC the transverse momentum distributions decrease
much faster. This fast fall-off limits the real measurements
to relatively small transverse momenta of the order of
10 GeV. The inclusion of realistic charge distribution is here
much more important than for the CMS conditions. The relative 
effect of damping with respect to the results with the monopole
charge form factor is shown in the right panel.
At $p_t$ = 10 GeV the damping factor is as big as 100!
Experiments at RHIC have a potential to confirm this prediction.

\begin{figure}[!h]    
\begin{minipage}[t]{0.46\textwidth}
\centering
\includegraphics[width=1\textwidth]{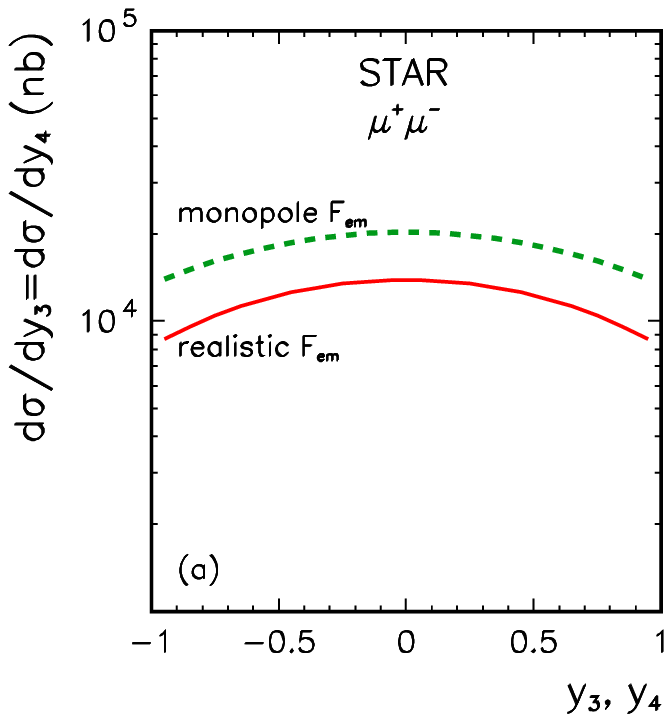}
\end{minipage}
\hspace{0.03\textwidth}
\begin{minipage}[t]{0.46\textwidth}
\centering
\includegraphics[width=1\textwidth]{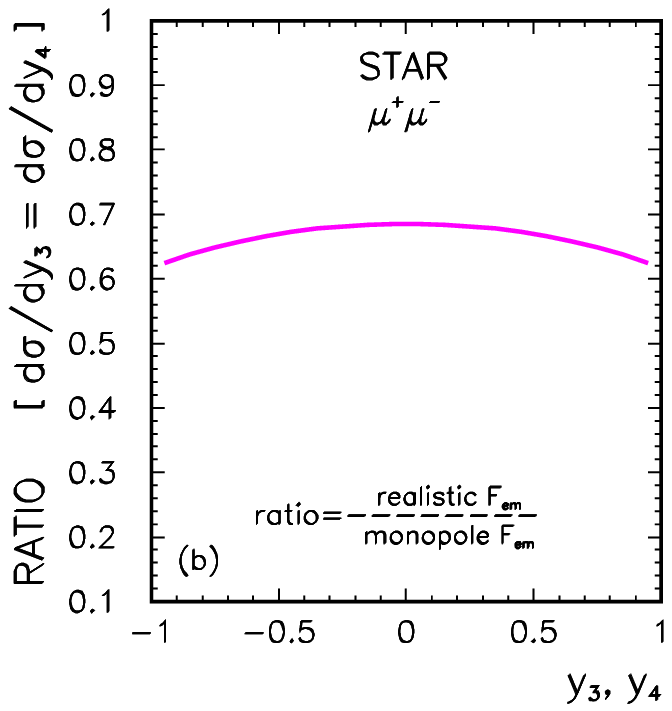}
\end{minipage}
   \caption{\label{fig:dsig_dy3_STAR}
   \small
(Color online) $\frac{{\rm d} \sigma}{{\rm d} y_3}$ (left) and the ratio (right) 
for the STAR conditions: $y_3, y_4 \in (-1,1)$, $p_{3t}, p_{4t} 
\geq$ 1 GeV and $W_{NN}=200$ GeV.
}
\end{figure}

In general, one could also inspect the rapidity distributions.
Our predictions are shown in 
Fig.\ref{fig:dsig_dy3_STAR}. We predict the 30-40 \% cross section damping
with respect to the reference calculation (monopole
charge form factor).

\begin{figure}[!h]    
\begin{minipage}[t]{0.46\textwidth}
\centering
\includegraphics[width=1\textwidth]{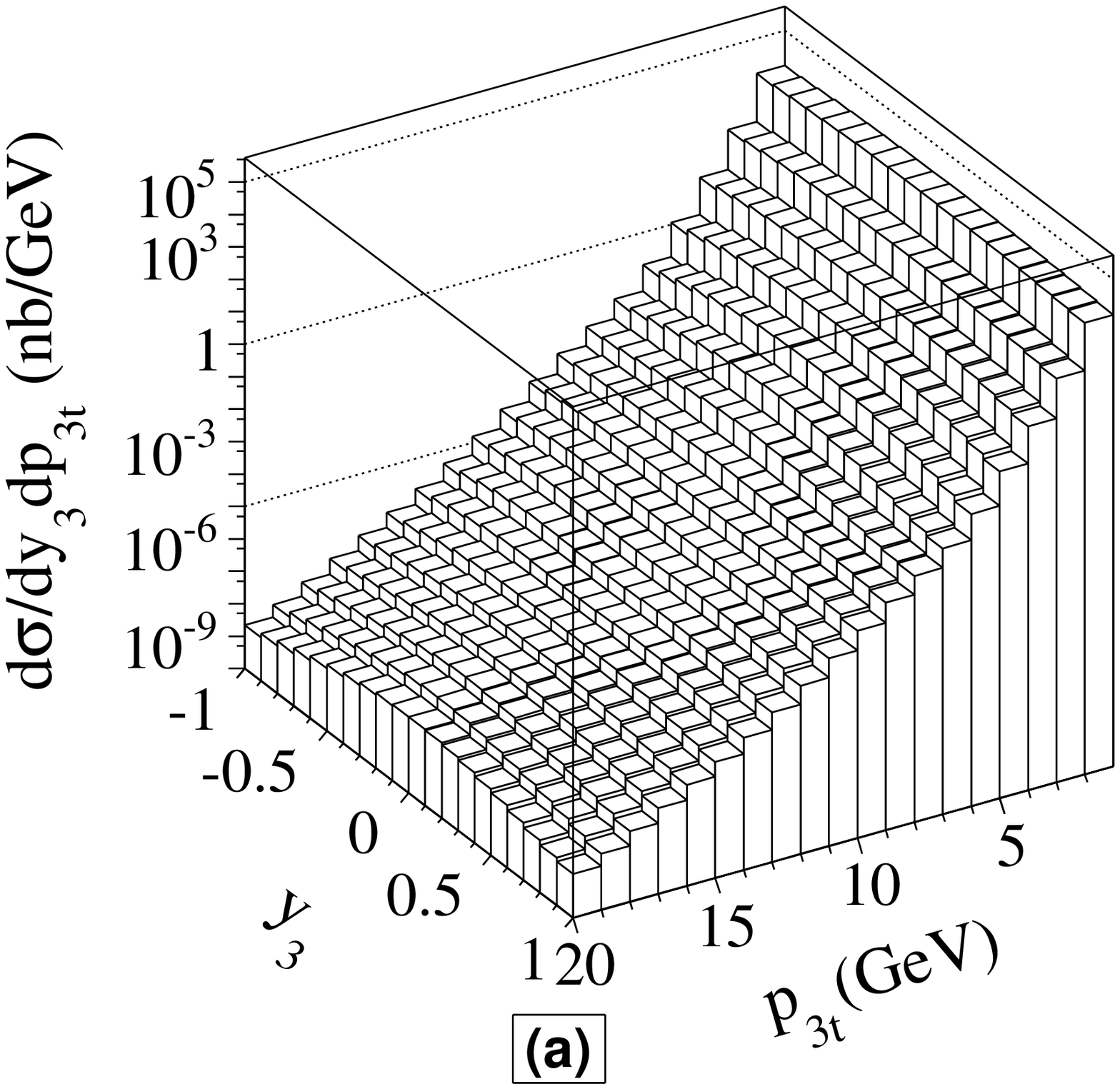}
\end{minipage}
\hspace{0.03\textwidth}
\begin{minipage}[t]{0.46\textwidth}
\centering
\includegraphics[width=1\textwidth]{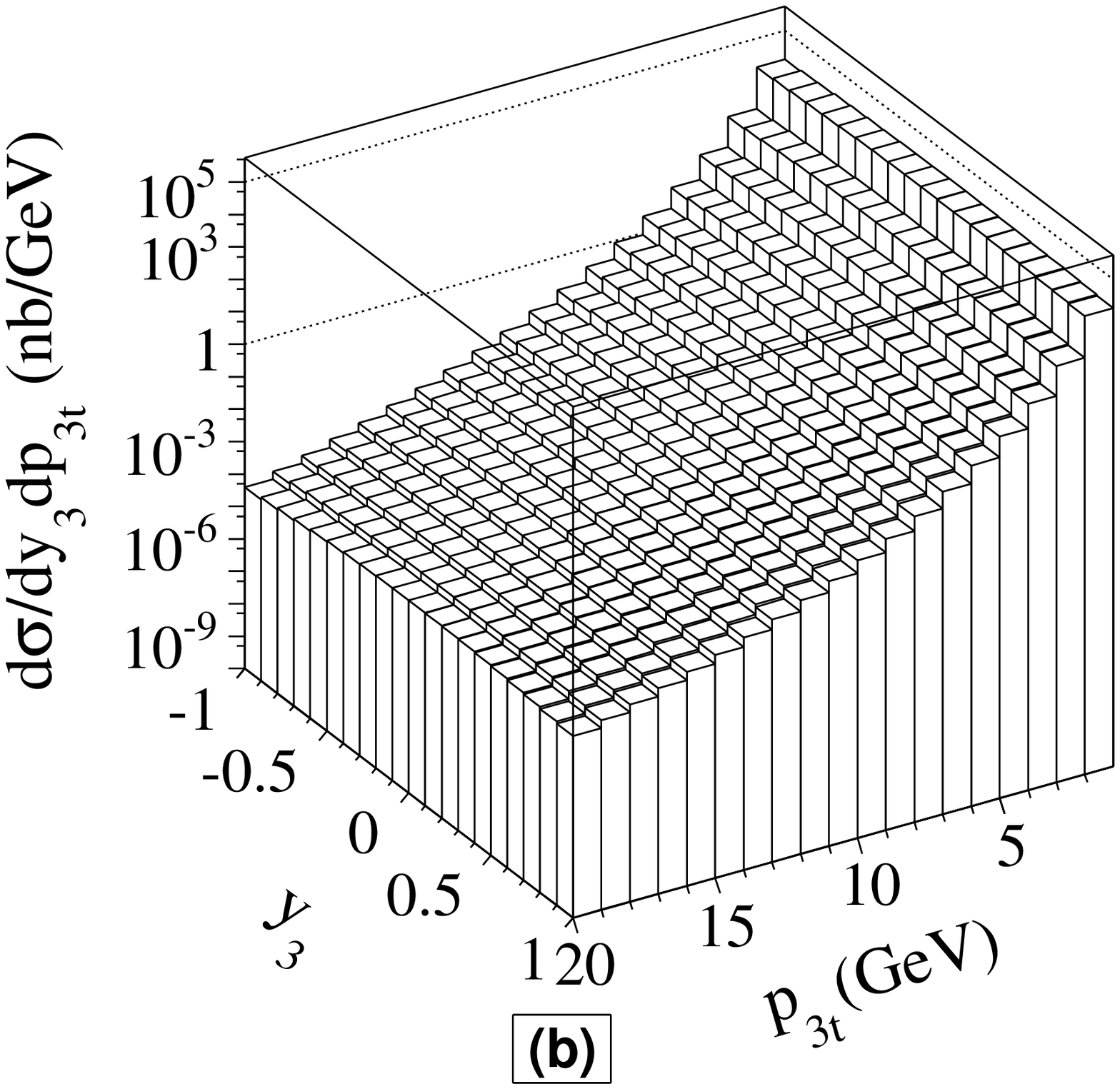}
\end{minipage}
   \caption{\label{fig:y3p3t_STAR}
   \small $\frac{{\rm d} \sigma}{{\rm d} y_3 \, {\rm d} p_{3t}}$ for realistic (left) and monopole (right) form factor for the STAR conditions $y_3, y_4 \in (-1,1)$, $p_{3t}, p_{4t} \geq 1$ GeV and $W_{NN}=200$ GeV. 
}
\end{figure}

\begin{figure}[!h]    
\begin{minipage}[t]{0.46\textwidth}
\centering
\includegraphics[width=1\textwidth]{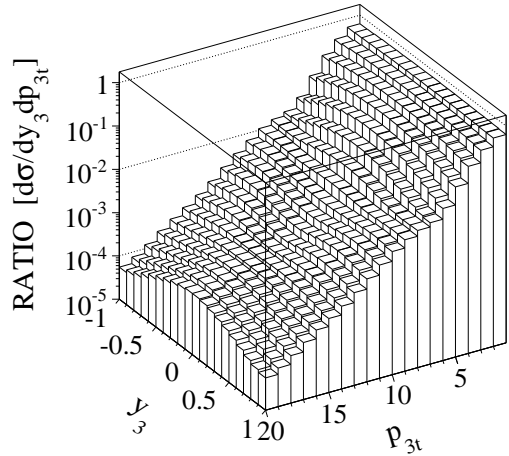}
\end{minipage}
   \caption{\label{fig:y3p3t_ratio_STAR}
   \small 
The ratio of the two-dimensional distributions from the previous figure
for the STAR conditions: $y_3, y_4 \in (-1,1)$, $p_{3t}, p_{4t} \geq 1$ GeV and $W_{NN}=200$ GeV. 
}
\end{figure}

\begin{figure}[!h]    
\begin{minipage}[t]{0.46\textwidth}
\centering
\includegraphics[width=1\textwidth]{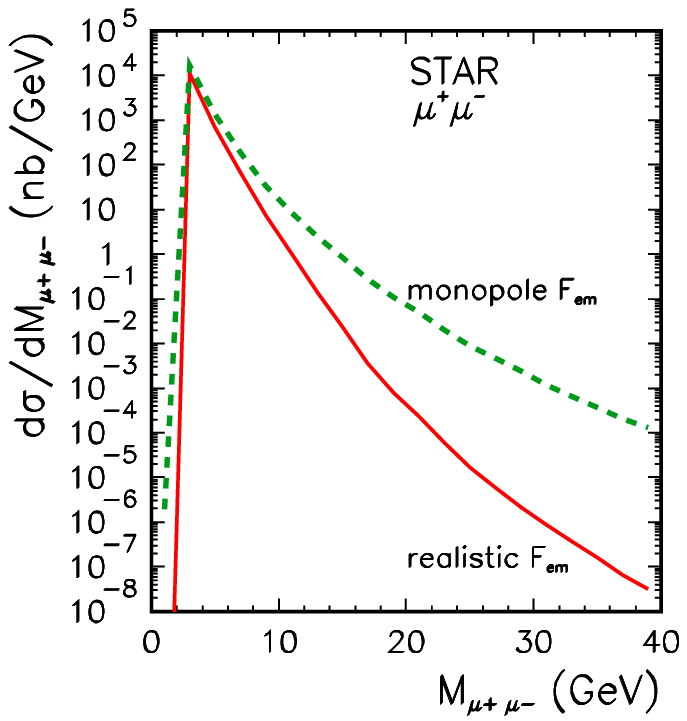}
\end{minipage}
   \caption{\label{fig:dsig_dm34_STAR}
   \small 
(Color online) Invariant mass distribution $\frac{{\rm d} \sigma}{{\rm d} M_{\mu^+ \mu^-}}$  
for the STAR conditions: $y_3, y_4 \in (-1,1)$, $p_{3t}, p_{4t} \geq 1$ GeV and $W_{NN}=200$ GeV. 
}
\end{figure}

The two-dimensional distributions in muon rapidity and
muon transverse momenta are shown in Fig.\ref{fig:y3p3t_STAR}
for the realistic and monopole form factors. Their
ratio is presented in Fig.\ref{fig:y3p3t_ratio_STAR}. Again as for
the transverse momentum distribution (see Fig.\ref{fig:dsig_dp3t_STAR}) 
a huge damping can be observed. The irregular structure of 
the ratio reflects the strong nonmonotonic dependence of 
the charge form factors of $Au$ nuclei on $t_1$ and $t_2$.
For completeness in Fig.\ref{fig:dsig_dm34_STAR} we show the
distribution of the dimuon invariant mass. The effect of the 
form factor oscillations shows up at large dimuon
invariant masses where the "realistic" cross section is
rather small.

\begin{figure}[!h]    
\begin{minipage}[t]{0.46\textwidth}
\centering
\includegraphics[width=1\textwidth]{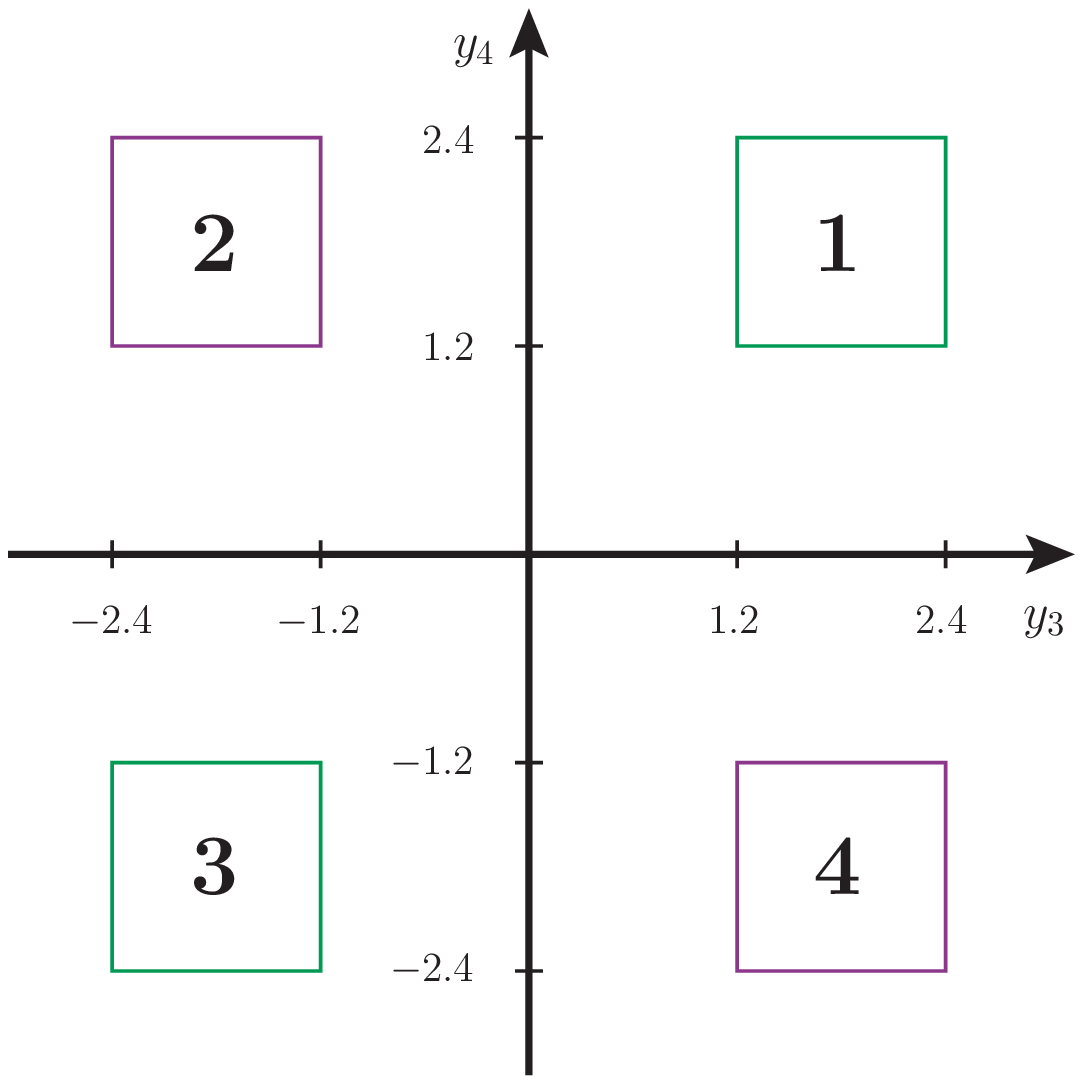}
\end{minipage}
\caption{ \label{fig:PHENIX_squares}
\small (Color online) The muon rapidity regions available by the PHENIX detector. 
}
\end{figure}

\begin{figure}[!h]    
\begin{minipage}[t]{0.46\textwidth}
\centering
\includegraphics[width=1\textwidth]{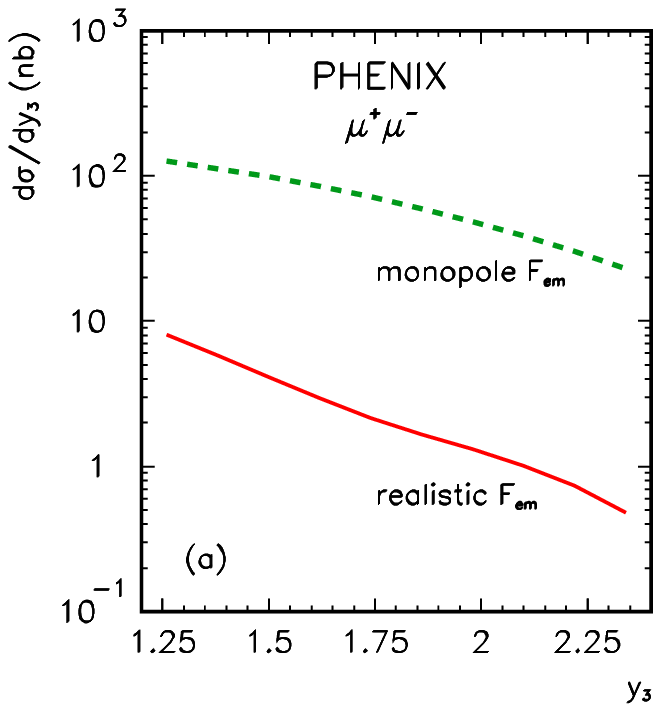}
\end{minipage}
\hspace{0.03\textwidth}
\begin{minipage}[t]{0.46\textwidth}
\centering
\includegraphics[width=1\textwidth]{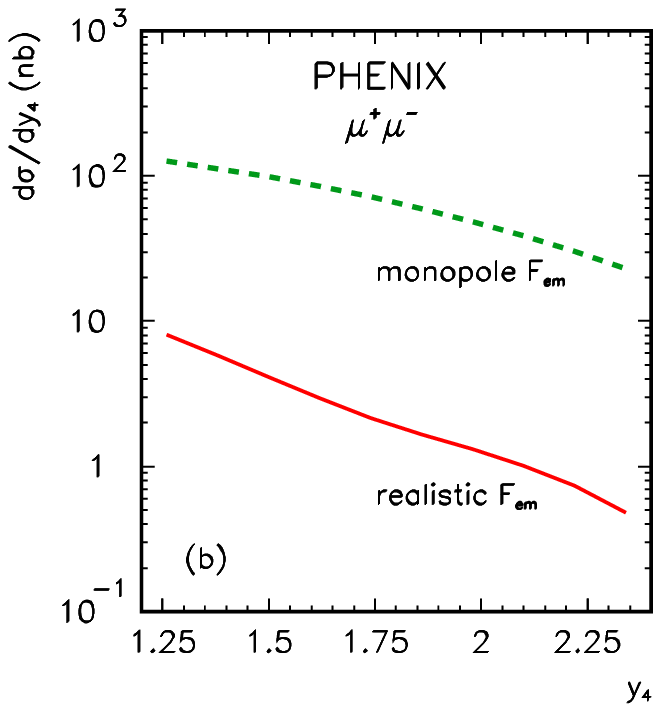}
\end{minipage}
   \caption{\label{fig:dsig_dy_1_PHENIX}
   \small  
(Color online) SQUARE 1: $\frac{{\rm d} \sigma}{{\rm d} y_3}$ (left) and 
$\frac{{\rm d} \sigma}{{\rm d} y_4}$ (right)
for the PHENIX conditions: 
1.2 $< |y_3, y_4 | <$ 2.4, $p_{3t}, p_{4t} \geq$ 2 GeV and $W_{NN}=200$ GeV.
}
\end{figure}

\begin{figure}[!h]    
\begin{minipage}[t]{0.46\textwidth}
\centering
\includegraphics[width=1\textwidth]{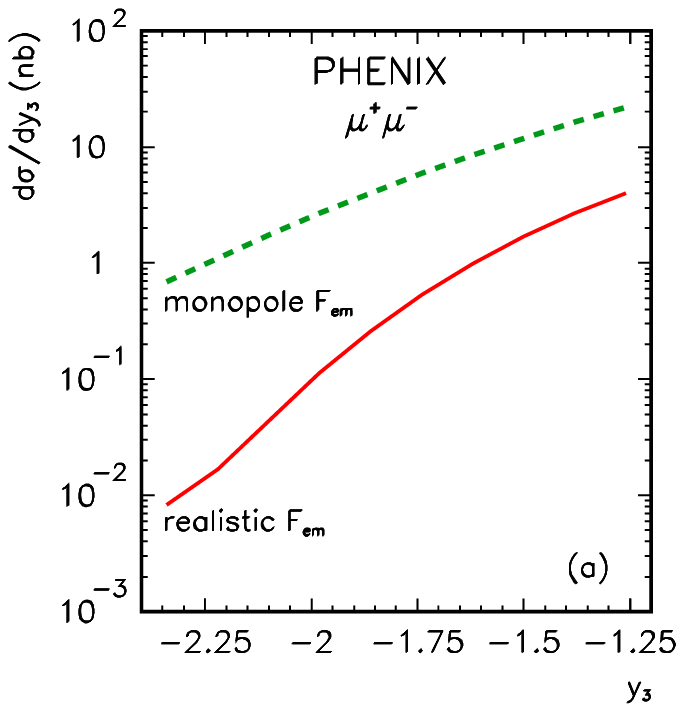}
\end{minipage}
\hspace{0.03\textwidth}
\begin{minipage}[t]{0.46\textwidth}
\centering
\includegraphics[width=1\textwidth]{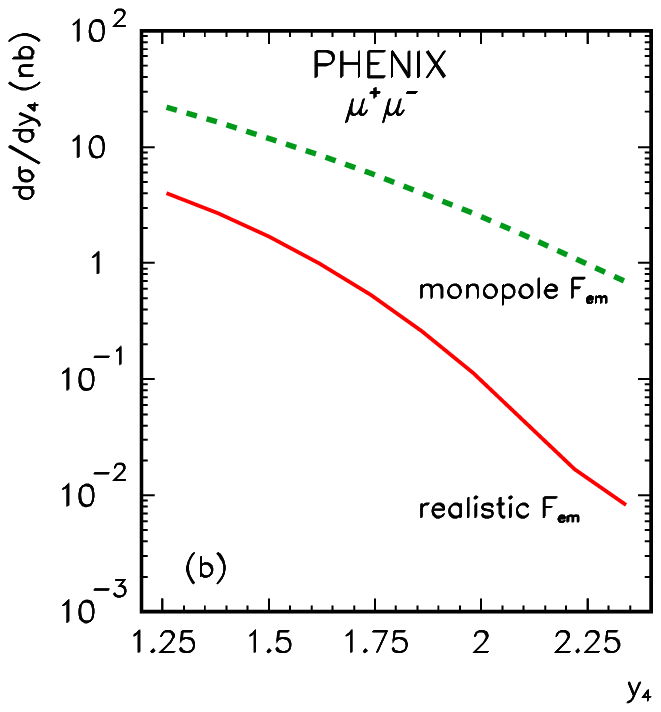}
\end{minipage}
   \caption{\label{fig:dsig_dy_2_PHENIX}
   \small 
   (Color online) SQUARE 2:  $\frac{{\rm d} \sigma}{{\rm d} y_3}$ (left) and
$\frac{{\rm d} \sigma}{{\rm d} y_4}$ (right)
for the PHENIX conditions: 
1.2 $ < |y_3, y_4 | < $ 2.4, $p_{3t}, p_{4t} \geq$ 2 GeV and $W_{NN}=200$ GeV.
}
\end{figure}

The PHENIX collaboration can measure muons in a rather limited 
range of rapidities shown in Fig.\ref{fig:PHENIX_squares}.
We have given names to the four possible regions (squares) in 
the figure.
In spite of these limitations, still interesting measurements 
can be done.
As an example in Fig.\ref{fig:dsig_dy_1_PHENIX} and 
Fig.\ref{fig:dsig_dy_2_PHENIX}
we show our predictions for SQUARE1 and SQUARE2, respectively
(the results for SQUARE3 and SQUARE4 are not shown as can 
be obtained by symmetry).
Again large deviations from the monopole form factor results are predicted.

\section{Conclusions}

The production of charge leptons in heavy ion collisions was 
proposed recently as a "laboratory" for studying Quantum Electrodynamics 
effects, in particular the multiple photon exchanges. 
While very interesting theoretically it is still nonrealistic because 
of other approximations made in the calculations. 

In this paper we have presented a study of the role of charge 
density for the differential distributions of muons produced 
in exclusive ultra-peripheral production in ultrarelativistic 
heavy ion collisions. 
Most of the calculations in the literature use so-called monopole 
charge form factor, which allows to write several formulae analytically. 
While it may be reasonable for the total rate of the dimuon production 
it is certainly too crude for differential distributions and 
for the cross sections with extra cuts imposed on transverse 
momenta of muons.

We have performed calculations in the Equivalent Photon Approximation in
the impact parameter space and in the momentum space using
Feynman diagrammatic approach. The first method is very convenient 
to include absorption effects, while the second one allows
to study differential distributions.

Our calculations show that the results obtained with the realistic and 
the approximate form factors can differ considerably, in some parts
of the phase space even by orders of magnitude.
The effects related to the charge distribution in nuclei are 
particularly important at large rapidities of muons and at 
large transverse momenta of muons.

We have also discussed the role of absorption effects which can be 
easily estimated in the impact parameter space. This allows 
to estimate the absorption effects for the total rate or for the 
rapidity distribution of the dimuon pairs. Estimating this effect 
in the case of differential distributions is not simple, but could be studied 
in the future.

We have presented predictions for the STAR and PHENIX detectors 
at RHIC as well as for the ALICE and CMS detectors at LHC.
In all cases we have found significant deviations from the reference
calculation for the monopole form factor.
It would be interesting to pin down the effects discussed here
and verify the present predictions in future studies at LHC. 
Both ALICE and CMS detectors could be used in such studies.

In practice such studies may not be simple as an efficient trigger
for the peripheral collisions is required. The multiphoton 
exchanges leading to additional excitation of nuclei and 
subsequent emission of neutrons could be useful in this context
(see e.g. \cite{BGKN09}). 
The neutrons could be then measured by the Zero Degree 
Calorimeters. First measurements of this type for $e^+ e^-$ pair emission
have been already performed by the STAR and PHENIX collaborations
\cite{STAR,PHENIX}. 

In the present calculation we have restricted
to lowest-order QED calculations paying a special attention
to realistic form factors and absorption effects
and totally ignored higher-order corrections.
How important are the QED higher-order correction was
demonstrated recently in Refs.\cite{Baltz2,SB}.
While Jentschura and Serbo \cite{SB} argue that
the higher-order corrections are rather small,
Baltz \cite{Baltz2} finds a huge reduction of the integrated 
cross section of the order of 20\%. Cleary the discrepancy
should be clarified in the future.
It would be also very interesting to calculate
the higher-order corrections for differential
distributions which will be measured at LHC
\footnote{All LHC experiments: ATLAS, CMS and ALICE,
will impose severe cuts on transverse momenta and
pseudorapidities of charged leptons.}.
The latter calculations seem to us rather difficult 
technically.

In the moment it seems precocious to answer the question whether 
the processes discussed here could be used as a luminosity monitor for heavy 
ion collisions at LHC.
In our opinion, first these processes should be measured and compared
to theoretical calculations.
In addition, the influence of the absorption effects and multiphoton
processes on differential distributions should be studied
in more detail.

\vspace{1cm}

{\bf Acknowledgments}

The discussion with Janusz Chwastowski, Jacek Oko{\l}owicz, Wolfgang Sch\"afer and 
Valeryi Serbo is acknowledged.
This work was partially supported by the Polish grant 
N N202 078735.


%
\end{document}